\documentclass[%
 amsmath,amssymb,
 aps,iop
]{emulateapj}

\usepackage{graphicx}
\usepackage{dcolumn}
\usepackage{bm}
\usepackage{float}
\usepackage{natbib}
\usepackage{multirow}
\usepackage[caption=false]{subfig}
\usepackage{lineno}
\usepackage[english, status=final]{fixme}
\fxusetheme{color}
\usepackage{color}
\usepackage{amsmath}
\usepackage{needspace}
\usepackage{array}
\usepackage{booktabs}
\newcolumntype{M}[1]{>{\centering\arraybackslash}m{#1}}
\newcommand\Tstrut{\rule{0pt}{2.9ex}}         
\newcommand\Bstrut{\rule[-1.2ex]{0pt}{0pt}}   
\newcommand\TBstrut{\Tstrut\Bstrut}           
\newcommand{\snana}{{{\tt SNANA}}}

\newcommand{\ler}{$4\times 10^{-3}$} 
\newcommand{\her}{1-2} 
\newcommand{\ligorate}{$1.26\times10^{4}$ yr$^{-1}$Gpc$^{-3}$}

\newcommand{\grbrateFong}{$90$ yr$^{-1}$Gpc$^{-3}$}
\newcommand{\grbrateFongFull}{$270^{+1580}_{-180}$ yr$^{-1}$Gpc$^{-3}$}
\newcommand{\diffimg}{{\tt DiffImg}}
\newcommand{\zphot}{$z_{\rm phot}$}
\newcommand{\diz}{$\delta_{iz}$}
\newcommand{\shallowTrigs}{3487} 
\newcommand{\deepTrigs}{1236} 
\newcommand{\maxKNratelim}{1.0$\times 10^{7}$ Gpc$^{-3}$yr$^{-1}$} 
\newcommand{\minKNratelim}{2.4$\times 10^{4}$ Gpc$^{-3}$yr$^{-1}$} 
\newcommand{\maxKNratelimd}{4.6$\times 10^{7} $ Gpc$^{-3}$yr$^{-1}$} 
\newcommand{\minKNratelimd}{5.2$\times 10^{4} $Gpc$^{-3}$yr$^{-1}$} 
\newcommand{\NCUTSDATA}{0}   
\newcommand{\NCUTSSIM}{$1.1\pm0.2$}  
\newcommand{\NPASSSND}{0.96}
\newcommand{\NPASSSNS}{0.17}
\newcommand{\NGENIa}{$3.7 \times 10^5$}
\newcommand{\NGENCC}{$4.3 \times 10^6$}
\newcommand{\mSB}{m_{\rm SB}}
\newcommand{\effKN}{\epsilon_{\rm KN}}
\newcommand{\NKN}{N_{\rm KN}}
\newcommand{\ZPmedian}{$\rm ZP_{median}$}
\newcommand{\shallowEcliptic}{42}
\newcommand{\shallowNoEcliptic}{11}
\newcommand{\deepEcliptic}{4}

\newcommand{\SNSIMSCALE}{40}
\newcommand{\IaRedshift}{1.1}
\newcommand{\CCRedshift}{0.8}

\newcommand{\dimmestpeakmagi}{$M_i = -11.4$}
\newcommand{\brightestpeakmagi}{$M_i = -16.2$}
\newcommand{\effdiffimg}{\epsilon_{\diffimg}}
\newcommand{\effsnana}{\epsilon_{\snana}}
\newcommand{\effnohost}{\epsilon_{{\rm No Host},\snana}}
\newcommand{\effnohostd}{\epsilon_{{\rm No Host},\diffimg}}
\newcommand{\effratio}{\mathcal{R}_\epsilon}
\newcommand{\KN}{{\small KN}}
\newcommand{\KNe}{{\small KN}e}
\newcommand{\LIGO}{{\small LIGO}}
\newcommand{\BK}{{\small BK}13}
\newcommand{\GW}{{\small GW}}
\newcommand{\EM}{{\small EM}}
\newcommand{\SN}{{\small SN}}
\newcommand{\SNe}{{\small SN}e}
\newcommand{\GRB}{{\small GRB}}
\newcommand{\SGRB}{{\small SGRB}}
\newcommand{\DESSN}{{\small DES}-{\small SN}}
\newcommand{\DES}{{\small DES}}
\newcommand{\BNS}{{\small BNS}}

\begin{document}

\preprint{APS/123-QED}

\submitted{Accepted to ApJ January 26, 2017}

\title{A Search for Kilonovae in the Dark Energy Survey}

\def\andname{}

\author{
Z.~Doctor\altaffilmark{1,2,3},
R.~Kessler\altaffilmark{1,4},
H.~Y.~Chen\altaffilmark{1,4},
B.~Farr\altaffilmark{1,2,5},
D.~A.~Finley\altaffilmark{6},
R.~J.~Foley\altaffilmark{7},
D.~A.~Goldstein\altaffilmark{8,9},
D.~E.~Holz\altaffilmark{1,2,4,5},
A.~G.~Kim\altaffilmark{9},
E.~Morganson\altaffilmark{10},
M.~Sako\altaffilmark{11},
D.~Scolnic\altaffilmark{1},
M.~Smith\altaffilmark{12},
M.~Soares-Santos\altaffilmark{6},
H.~Spinka\altaffilmark{13},
T. M. C.~Abbott\altaffilmark{14},
F.~B.~Abdalla\altaffilmark{15,16},
S.~Allam\altaffilmark{6},
J.~Annis\altaffilmark{6},
K.~Bechtol\altaffilmark{17},
A.~Benoit-L{\'e}vy\altaffilmark{18,15,19},
E.~Bertin\altaffilmark{18,19},
D.~Brooks\altaffilmark{15},
E.~Buckley-Geer\altaffilmark{6},
D.~L.~Burke\altaffilmark{20,21},
A. Carnero Rosell\altaffilmark{22,23},
M.~Carrasco~Kind\altaffilmark{24,10},
J.~Carretero\altaffilmark{25,26},
C.~E.~Cunha\altaffilmark{20},
C.~B.~D'Andrea\altaffilmark{27,12},
L.~N.~da Costa\altaffilmark{22,23},
D.~L.~DePoy\altaffilmark{28},
S.~Desai\altaffilmark{29},
H.~T.~Diehl\altaffilmark{6},
A.~Drlica-Wagner\altaffilmark{6},
T.~F.~Eifler\altaffilmark{30},
J.~Frieman\altaffilmark{6,1},
J.~Garc\'ia-Bellido\altaffilmark{31},
E.~Gaztanaga\altaffilmark{25},
D.~W.~Gerdes\altaffilmark{32},
R.~A.~Gruendl\altaffilmark{24,10},
J.~Gschwend\altaffilmark{22,23},
G.~Gutierrez\altaffilmark{6},
D.~J.~James\altaffilmark{33,14},
E.~Krause\altaffilmark{20},
K.~Kuehn\altaffilmark{34},
N.~Kuropatkin\altaffilmark{6},
O.~Lahav\altaffilmark{15},
T.~S.~Li\altaffilmark{6,28},
M.~Lima\altaffilmark{35,22},
M.~A.~G.~Maia\altaffilmark{22,23},
M.~March\altaffilmark{11},
J.~L.~Marshall\altaffilmark{28},
F.~Menanteau\altaffilmark{24,10},
R.~Miquel\altaffilmark{36,26},
E.~Neilsen\altaffilmark{6},
R.~C.~Nichol\altaffilmark{27},
B.~Nord\altaffilmark{6},
A.~A.~Plazas\altaffilmark{30},
A.~K.~Romer\altaffilmark{37},
E.~Sanchez\altaffilmark{38},
V.~Scarpine\altaffilmark{6},
M.~Schubnell\altaffilmark{32},
I.~Sevilla-Noarbe\altaffilmark{38},
R.~C.~Smith\altaffilmark{14},
F.~Sobreira\altaffilmark{22,39},
E.~Suchyta\altaffilmark{40},
M.~E.~C.~Swanson\altaffilmark{10},
G.~Tarle\altaffilmark{32},
A.~R.~Walker\altaffilmark{14},
W.~Wester\altaffilmark{6}
\\ \vspace{0.2cm} (DES Collaboration) \\
}
 
\affil{$^{1}$ Kavli Institute for Cosmological Physics, University of Chicago, Chicago, IL 60637, USA}
\affil{$^{2}$ Department of Physics, University of Chicago, Chicago, IL 60637, USA}
\affil{$^{3}$ NSF GRFP Fellow}
\affil{$^{4}$ Department of Astronomy and Astrophysics, University of Chicago, Chicago, IL 60637, USA}
\affil{$^{5}$ Enrico Fermi Institute, University of Chicago, Chicago, IL 60637, USA}
\affil{$^{6}$ Fermi National Accelerator Laboratory, P. O. Box 500, Batavia, IL 60510, USA}
\affil{$^{7}$ Department of Astronomy and Astrophysics, University of California, Santa Cruz, CA 95064, USA}
\affil{$^{8}$ Department of Astronomy, University of California, Berkeley,  501 Campbell Hall, Berkeley, CA 94720, USA}
\affil{$^{9}$ Lawrence Berkeley National Laboratory, 1 Cyclotron Road, Berkeley, CA 94720, USA}
\affil{$^{10}$ National Center for Supercomputing Applications, 1205 West Clark St., Urbana, IL 61801, USA}
\affil{$^{11}$ Department of Physics and Astronomy, University of Pennsylvania, Philadelphia, PA 19104, USA}
\affil{$^{12}$ School of Physics and Astronomy, University of Southampton,  Southampton, SO17 1BJ, UK}
\affil{$^{13}$ Argonne National Laboratory, 9700 South Cass Avenue, Lemont, IL 60439, USA}
\affil{$^{14}$ Cerro Tololo Inter-American Observatory, National Optical Astronomy Observatory, Casilla 603, La Serena, Chile}
\affil{$^{15}$ Department of Physics \& Astronomy, University College London, Gower Street, London, WC1E 6BT, UK}
\affil{$^{16}$ Department of Physics and Electronics, Rhodes University, PO Box 94, Grahamstown, 6140, South Africa}
\affil{$^{17}$ LSST, 933 North Cherry Avenue, Tucson, AZ 85721, USA}
\affil{$^{18}$ CNRS, UMR 7095, Institut d'Astrophysique de Paris, F-75014, Paris, France}
\affil{$^{19}$ Sorbonne Universit\'es, UPMC Univ Paris 06, UMR 7095, Institut d'Astrophysique de Paris, F-75014, Paris, France}
\affil{$^{20}$ Kavli Institute for Particle Astrophysics \& Cosmology, P. O. Box 2450, Stanford University, Stanford, CA 94305, USA}
\affil{$^{21}$ SLAC National Accelerator Laboratory, Menlo Park, CA 94025, USA}
\affil{$^{22}$ Laborat\'orio Interinstitucional de e-Astronomia - LIneA, Rua Gal. Jos\'e Cristino 77, Rio de Janeiro, RJ - 20921-400, Brazil}
\affil{$^{23}$ Observat\'orio Nacional, Rua Gal. Jos\'e Cristino 77, Rio de Janeiro, RJ - 20921-400, Brazil}
\affil{$^{24}$ Department of Astronomy, University of Illinois, 1002 W. Green Street, Urbana, IL 61801, USA}
\affil{$^{25}$ Institut de Ci\`encies de l'Espai, IEEC-CSIC, Campus UAB, Carrer de Can Magrans, s/n,  08193 Bellaterra, Barcelona, Spain}
\affil{$^{26}$ Institut de F\'{\i}sica d'Altes Energies (IFAE), The Barcelona Institute of Science and Technology, Campus UAB, 08193 Bellaterra (Barcelona) Spain}
\affil{$^{27}$ Institute of Cosmology \& Gravitation, University of Portsmouth, Portsmouth, PO1 3FX, UK}
\affil{$^{28}$ George P. and Cynthia Woods Mitchell Institute for Fundamental Physics and Astronomy, and Department of Physics and Astronomy, Texas A\&M University, College Station, TX 77843,  USA}
\affil{$^{29}$ Department of Physics, IIT Hyderabad, Kandi, Telangana 502285, India}
\affil{$^{30}$ Jet Propulsion Laboratory, California Institute of Technology, 4800 Oak Grove Dr., Pasadena, CA 91109, USA}
\affil{$^{31}$ Instituto de F\'{\i}sica Te\'orica UAM-CSIC, Universidad Auton\'oma de Madrid, Cantoblanco, 28049 Madrid, Spain}
\affil{$^{32}$ Department of Physics, University of Michigan, Ann Arbor, MI 48109, USA}
\affil{$^{33}$ Astronomy Department, University of Washington, Box 351580, Seattle, WA 98195, USA}
\affil{$^{34}$ Australian Astronomical Observatory, North Ryde, NSW 2113, Australia}
\affil{$^{35}$ Departamento de F\'{\i}sica Matem\'atica,  Instituto de F\'{\i}sica, Universidade de S\~ao Paulo,  CP 66318, CEP 05314-970, S\~ao Paulo, SP,  Brazil}
\affil{$^{36}$ Instituci\'o Catalana de Recerca i Estudis Avan\c{c}ats, E-08010 Barcelona, Spain}
\affil{$^{37}$ Department of Physics and Astronomy, Pevensey Building, University of Sussex, Brighton, BN1 9QH, UK}
\affil{$^{38}$ Centro de Investigaciones Energ\'eticas, Medioambientales y Tecnol\'ogicas (CIEMAT), Madrid, Spain}
\affil{$^{39}$ Universidade Federal do ABC, Centro de Ci\^encias Naturais e Humanas, Av. dos Estados, 5001, Santo Andr\'e, SP, Brazil, 09210-580}
\affil{$^{40}$ Computer Science and Mathematics Division, Oak Ridge National Laboratory, Oak Ridge, TN 37831}

\date{\today}

\begin{abstract}
The coalescence of a binary neutron star (\BNS{}) pair is expected to produce gravitational waves (\GW{}) and electromagnetic (\EM{}) radiation, both of which may be detectable with currently available instruments. We describe a search for a predicted r-process optical transient from these mergers, dubbed the ``kilonova" (\KN{}), using $griz$ broadband data from the Dark Energy Survey Supernova Program (\DESSN{}).  Some models predict \KNe{} to be redder, shorter-lived, and dimmer than supernovae (\SNe{}), but the event rate of \KNe{} is poorly constrained.  We simulate \KN{} and \SN{} light curves with the Monte-Carlo simulation code \snana{} to optimize selection requirements, determine search efficiency, and predict \SN{} backgrounds. Our analysis of the first two seasons of \DESSN{} data results in \NCUTSDATA\ events, and is consistent with our prediction of \NCUTSSIM\ background events based on simulations of \SNe{}. From our prediction, there is a 33\% chance of finding 0 events in the data. Assuming no underlying galaxy flux, our search sets 90\% upper limits on the \KN{} volumetric rate of \maxKNratelim{} for the dimmest \KN{} model we consider (peak $i$-band absolute magnitude \dimmestpeakmagi{} mag) and \minKNratelim{} for the brightest (\brightestpeakmagi{} mag).  Accounting for anomalous subtraction artifacts on bright galaxies, these limits are $\sim 3$ times higher.   This analysis is the first untriggered optical \KN{} search and informs selection requirements and strategies for future \KN{} searches. Our upper limits on the \KN{} rate are consistent with those measured by \GW{} and gamma-ray burst searches.

\end{abstract}

\pacs{Valid PACS appear here}
\maketitle

\section{Introduction}
The recent detections by \LIGO{} of gravitational waves (\GW{}) from binary black hole mergers \citep{LIGO150914,GW151226} have motivated searches for electromagnetic (\EM{}) counterparts to gravitational waves \citep{LIGOEM,Annis,SoaresSantos,CowperthwaiteFollow}.  Theoretical and numerical studies suggest that outflows of energetic neutron-rich material during a binary neutron star (\BNS{}) merger enable r-process nucleosynthesis (e.g. \citealt{Li}).  The decay of these r-process elements results in isotropic thermal emission and is called a ``kilonova" (\KN{}).  \citet{Metzger:2011bv} compared different \EM{} counterparts of \GW{} sources and concluded that the kilonova has promising detectability with current instruments. Observations of \KNe{} could constrain models of neutron star mergers, and an accurate redshift measurement would allow \GW{} measurements to be used as cosmological distance probes \citep{Schutz,Dalal}. While properties of optical \KN{} light curves remain uncertain, a number of models predict \KNe{} to be dim ($M_i \sim-14$ mag), red ($i-z \sim1$ mag), and short-lived ($\sim$1 week) \citep{BK2013,Tanaka,redorblue,Lippuner}.  

An interesting \KN{} candidate was reported by \citet{Tanvir} and \citet{Berger:2013wna}.  The candidate was first identified by the gamma ray burst \GRB{}130613B, which triggered both the {\it Swift} Burst Alert Telescope and Konus-Wind.  Following \GRB{}130613B, the teams reported two epochs of Hubble Space Telescope observations in the V and H bands as well as optical observations with the Inamori Magellan Areal Camera and Spectrograph ({\small IMACS}) and the Low Dispersion Survey Spectrograph (LDSS3) on the Magellan Telescopes.  At 9.4 days after the \GRB{}, the source was found to have $M_H \sim -15.2$ mag and $M_V \gtrsim -13.3$ mag, indicating a red $V-H$ color $\gtrsim 1.9$ mag. Other r-process events possibly involving a neutron star have been reported in \citet{Jin}, \citet{Jin2016}, and \citet{Ji}.

Here we describe an independent search for \KNe{} from the first two seasons of data from the Dark Energy Survey supernova program (\DESSN{}) \citep{Bernstein,Diehl} using the Dark Energy Camera ({\small DEC}am: \citealt{Flaugher}).  \KN{} searches triggered by \GRB{}s or \GW{}s will be an integral part of \GW{} multi-messenger astronomy, but here we use existing \DESSN{} data to search for \KNe{} without an external trigger. With four optical broadband filters, 30-square-degree coverage in the supernova fields, and $\sim$1 week cadence in each filter with excellent depth per visit, the \DESSN{} sample is well-suited for a \KN{} search.   Light curve simulations of \KNe{} and supernova (\SN{}) backgrounds are used to inform the analysis and selection criteria that we apply to the \DESSN{} data sample to look for \KNe{}. The \KN{} simulations are based on spectral energy distributions (SED) resulting from radiation transport calculations of \KN{}-merger models from \citet{BK2013} (hereafter \BK{}). These calculations fold in r-process element opacities using atomic data for heavy elements rather than approximating their opacities with that of iron. As a result, \BK{} predict \KN{} light curves that are dimmer, redder, and longer-lived than those predicted with iron opacities. These models are still highly uncertain, so our analysis reports volumetric rate limits for each \KN{} model and over a wide range of absolute brightnesses.

We make a preliminary estimate of our potential \KN{} sensitivity using two approximate calculations based on previously published results. Both approximations assume a \KN{} peak $i$-band magnitude of $M_i \sim -14$ mag and 100\% search efficiency within the \DESSN{} limiting magnitudes and sky areas described in \S~\ref{sec:datasample}.  The first approximation assumes an optimistically large \KN{} rate given by the \LIGO{} 90\% confidence upper limit on binary neutron star mergers of \ligorate{} \citep{LIGO}.  With these assumptions, we would expect to find \her{} \KNe{} in our \DESSN{} sample.  The second calculation is based on the hypothetical correspondence between \KNe{} and short-hard gamma-ray bursts (\SGRB{}, \citealt{Paczynski1986}, \citealt{Narayan}). Assuming that the true event rate of \SGRB{}s is the lower limit found in \citet{FongGRB} of \grbrateFong{} and that $1/2$ of \SGRB{}s are associated with \KNe{}, the probability of seeing a \KN{} in our sample is $\sim$\ler{}. 

We present results from the first untriggered optical search for \KNe{}, which is complementary to the \LIGO{} search for \BNS{} mergers based on predicted \GW{} signals.  \LIGO{} directly probes for mergers by looking for characteristic ``chirp" \GW{} signals from sources in its detection volume \citep{LIGO}.  The \DES{} \KN{} search described here is sensitive to optical emission from such mergers.   While we do not discuss optical follow-up to \LIGO{} triggers, the methods and results presented here will inform strategies and contaminant rejection techniques for follow-up \citep{SoaresSantos}. The outline of this paper is as follows.  In \S~\ref{sec:datasample} and \S~\ref{sec:simulations}, we describe the \DES{} data sample and the simulations, respectively.  \S~\ref{sec:analysis} details selection requirements and our \KN{} search efficiency for different models of host galaxy noise.  The results and discussion of the analysis are presented in \S~\ref{sec:results} and \S~\ref{sec:discussion}, and we conclude in \S~\ref{sec:conclusion}.

\section{DES-SN Data Sample} 
\label{sec:datasample}
For \DESSN{}, the {\small CTIO} Blanco-4m telescope and {\small DEC}am were used to make repeated observations of ten 3 deg$^2$ fields.  Each field was observed in {\it griz} bands with central wavelengths of 4830, 6430, 7830, 9180~\AA, respectively. Eight of these fields were ``shallow" fields with an average single-visit depth of $\sim 23.5$ mag in each band. The other two were ``deep" fields with an average single-visit depth of $\sim 24.5$ mag in each band.  The exposure time and number of exposures per visit are given in Table 1 of \citet{DiffImg} (hereafter K15). The cadence was approximately one visit per week in each band and field.   Transients were detected using the difference-imaging pipeline \diffimg{} described in K15 to find events with signal-to-noise ratio above 5 that also pass automated scanning to reject subtraction artifacts \citep{Goldstein}. To select candidates for \SN{} science, \DESSN{} requires a detection on two separate nights, mainly to reject asteroids\footnote{See sections 3.2.4 and 3.3 in K15 for more details about science candidates}. However, {\it all} candidates with two detections are saved, even if the two detections are on the same night.  

For our \KN{} search, we consider events with $i$ and $z$ detections close in time, since models predict that \KNe{} will have higher SNR in the \DES{} $i$ and $z$ bands than in $g$ and $r$. Additionally, \KNe{} are expected to fade on the timescale of repeat \SN{}-field observations ($\sim1$ week). In the shallow fields, we require an $i$ and $z$ detection on the same night since all four ({\it griz}) bands are observed within $\sim~20$ minutes anyway. In the deep fields, adjacent-night $i$ and $z$ detections are accepted since the four bands are not always observed on the same night.  We define a {\it trigger} to be the first paired $i$ and $z$ detections of a candidate\footnote{Pairing requires that the RA and DEC of the $i$ and $z$ detections are within 1", and thus the angular (radial) separation extends out to $\sim\sqrt{2}\times1$".}. Longer time separations between $i$ and $z$-band detections are not considered as triggers since \KNe{} are expected to dim significantly between repeated observations of the fields. 

Using the first two seasons of \DESSN{}, these criteria result in \shallowTrigs{} triggers in the shallow fields, and \deepTrigs{} triggers in the deep fields, most of which are SNe and asteroids.  Our data sample includes PSF-fitted measurements of the flux and its uncertainty in all four bands and all epochs, the angular separation between $i$ and $z$-band detections on the trigger night (for asteroid rejection), the matched host galaxy, and its photometric redshift.  The absolute photometric calibration of the sample has been validated at the 2\% level (K15).

\section{Simulations of DES Light Curves}
\label{sec:simulations}
Monte Carlo simulations of \KN{} and \SN{} light curves, using the \snana{} software package \citep{SNANA}, are employed to tune selection criteria, to determine the \KN{} search efficiency, and to predict backgrounds.  For an arbitrary light curve model, the \snana{} simulation uses the observing conditions (PSF, zero point, sky noise) at each \DESSN{} epoch and passband to generate a redshifted\footnote{The \snana{} simulations use a FLRW cosmology with $H_0=70$km/s/Mpc, $\Omega_M=0.3$, and $\Omega_{\Lambda}=0.7$.} flux and uncertainty that would have been measured by \diffimg. The \snana{} simulation generates light curves at the catalog-level and does not use images. As described in section 7 of K15, \snana{} also does not fully account for image subtraction artifacts on bright galaxies.  To better characterize the \KN{} efficiency, we perform an additional study of fake point sources overlaid on the search images near low-redshift galaxies and processed with \diffimg.  
 
\paragraph*{Kilonovae:}
\begin{figure*}[ht!]
	\centering
	\includegraphics[scale=0.45]{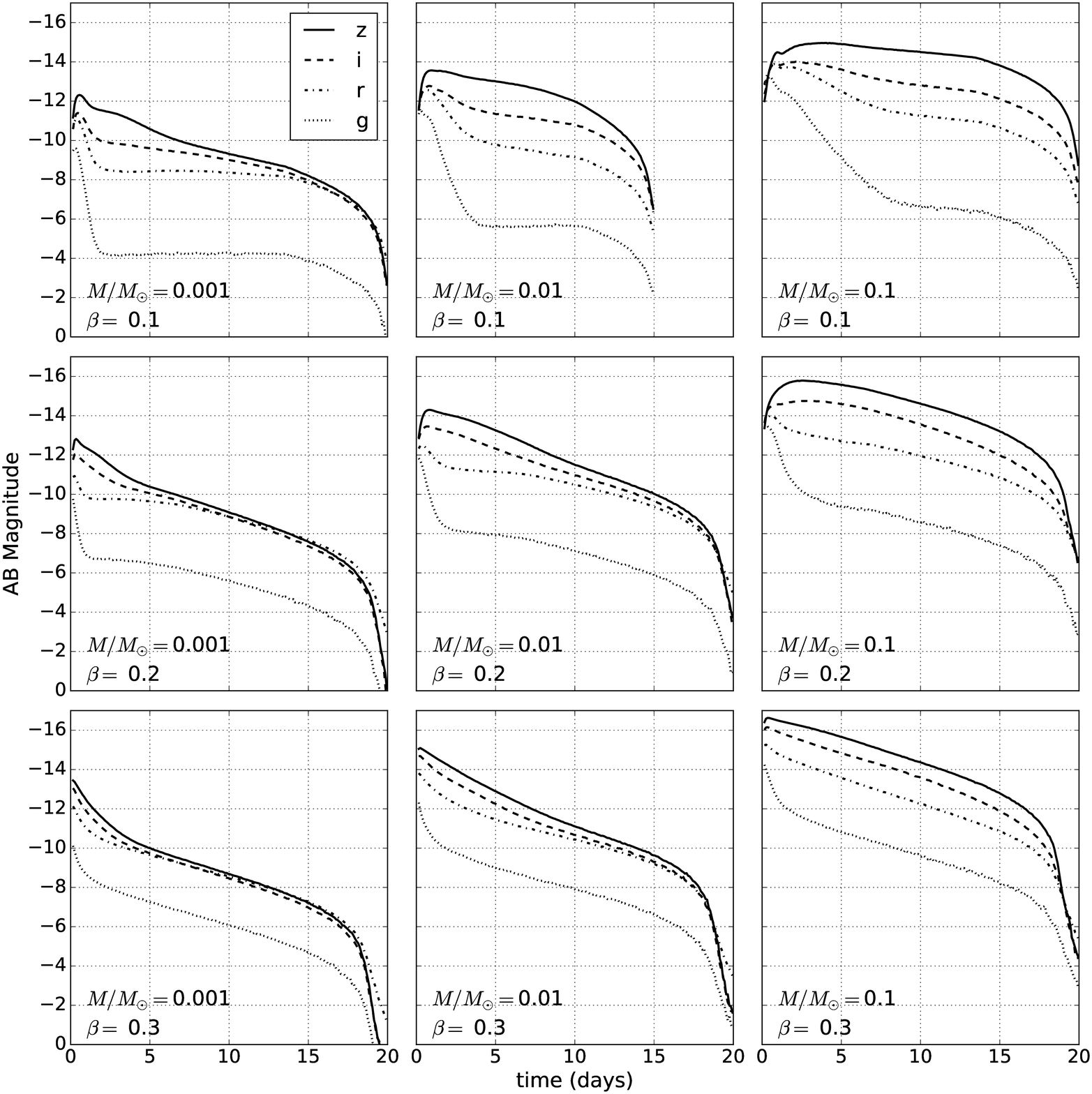}
	\caption{Computed $griz$ broadband light curves from integrating the nine BK13 spectral energy distributions. The bottom left of each panel shows the BK13 model parameters.}
	\label{fig:Magplot_allmodels}
\end{figure*}
We simulate \KN{} light curves using rest-frame SED models from \BK{}. Their models have evolved since the start of this analysis \citep{Barnes2016}, but here we restrict ourselves to the original light curves of \citet{BK2013}. The implications of the new models are discussed in \S~\ref{sec:conclusion}. The \BK{} models are parameterized by the velocity $\beta=v/c$ and mass $M$ of the matter ejected from the merger.  \BK{} generate nine SEDs corresponding to $\beta$ = [0.1, 0.2, 0.3] and $M$ = [0.001, 0.01, 0.1] $M_{\odot}$, which we use in our analysis. 
Calculated {\it griz} light curves for all nine models are shown in Fig~\ref{fig:Magplot_allmodels}. These light curves are generically brightest in the $z$ band, with decreasing brightness in bluer filters.  At the time of maximum $z$-flux, the ratio of $g$, $r$, and $i$-band flux to $z$-band flux is 0.04, 0.2, and 0.5, respectively, when averaged over all nine models.  Notably, the rest-frame brightness of these models spans 4-5 magnitudes, and the distribution over model parameters is unknown, adding to the uncertainty in overall \KN{} detectability.  In the \snana{} simulation, \KN{} light curve start times are randomly generated during the first season (Y1: 56534 $<$ MJD $<$ 56698) and second season (Y2: 56877 $<$ MJD $<$ 57067), and the redshift distribution is assumed to follow a constant co-moving volumetric rate. Our analysis considers \KN{} light curves with and without host galaxy flux.  When including host galaxy flux, the light curves are generated in the simulation such that the \KN{} rate at a particular location is proportional to the background host brightness as described in \S~\ref{subsec:hostgal}. To compute our sensitivity over a wide range of \KN{} brightness, absolute magnitude offsets are applied as described in \S~\ref{sec:analysis} and \S~\ref{sec:results}. 

  \begin{figure*}
	\centering
	\includegraphics[scale=0.55]{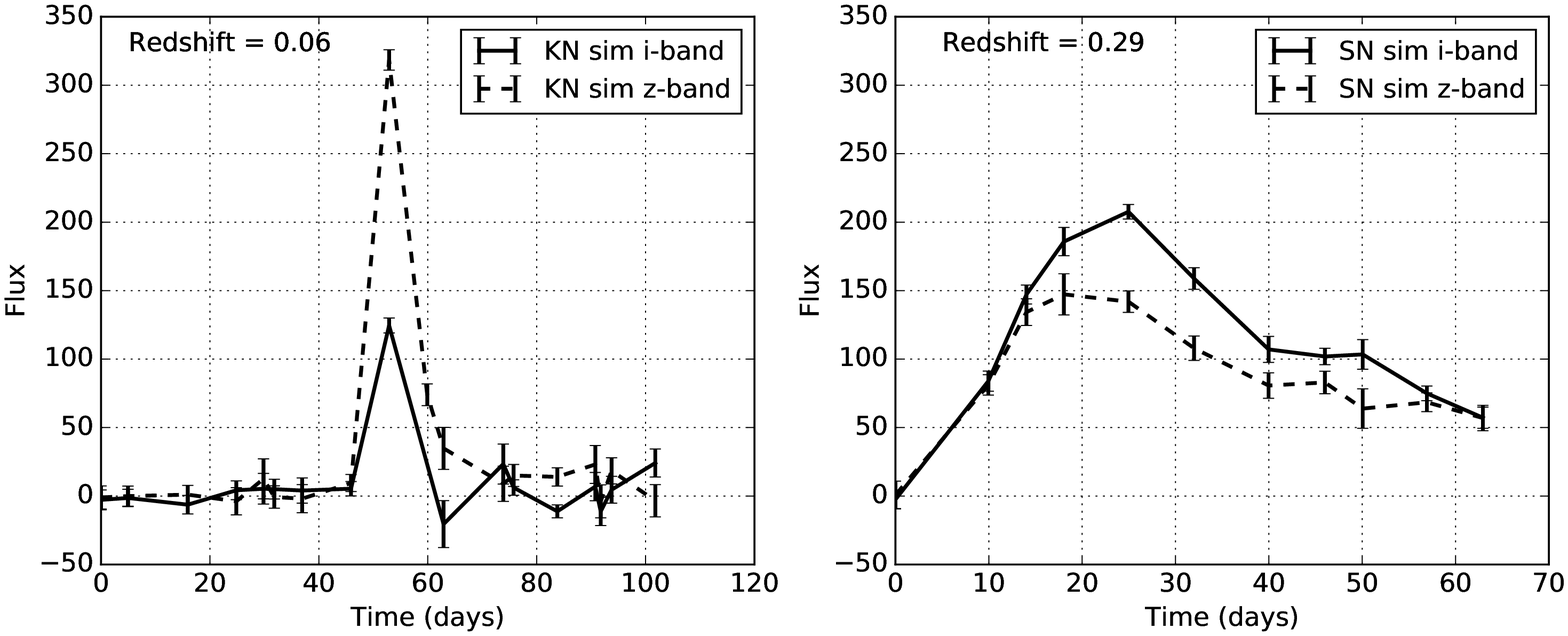}
	\caption{Observed \KN{} and \SN{} light curves in the $i$ and $z$ filters, as simulated with \snana{}.  The \KN{} is based on the \BK{} model with $\beta = 0.3$, $M = 0.1 M_{\odot}$ and redshift $z = 0.06$.  The SNIa is simulated with SALT-II color $c = 0.03$, stretch parameter $x_1 = -0.75$, and redshift $z = 0.29$. Magnitudes are given by $27.5 - 2.5\log_{10}({\rm Flux})$; e.g., the shallow-field detection limit of mag=23.5 corresponds to Flux=40. The error bars show the simulated flux and uncertainties for each observation; the lines connect these simulated points to guide the eye.}
	\label{fig:KNSNlc}
\end{figure*}

 \paragraph*{Supernovae:}
We perform simulations of Type Ia \SNe{} using the SALT-II light curve model \citep{SALTII}, the volumetric rate vs. redshift from ~\citet{Dilday}, and the `G10' intrinsic scatter model and the stretch and color populations from \citet{K13}. For core collapse (CC) \SNe{} we use CC templates based on IIbc, II-P, and IIn \SNe{} as described in \cite{K10} along with the rates from \citet{LiCC}.  Example $i$ and $z$-band light curves simulated for a typical \KN{} and \SN{}{\small I}a are shown in Fig.~\ref{fig:KNSNlc}.  In general, the \SNe{} tend to be bluer and have longer timescales than \KNe{}.  
\SN{} times of peak flux are simulated in the ranges MJD = 56450 to 56740 for Y1 and MJD = 56840 to 57110 for Y2.  Each window extends $\sim$ 2 months before the start of the observing season, and $\sim$ 1 month after the end. These extended time windows account for long rise and fall times of \SN{} light curves as well as the $(1+z)$ time dilation for higher redshift \SNe{}.  \SNe{} are generated in the redshift range $z<1.35$. The probability of matching to a host galaxy is taken from \citet{Bernstein}\footnote{See $m_i<24$ and $\kappa_{Ia}=0.5$ column of Table 18.}.  Note that the simulation does not account for \SNe{} matched to incorrect hosts.  For each simulated host galaxy, the associated photometric redshift (\zphot) is based on the photometric redshift distribution in the Science Verification (SV) catalogue described in \citet{Bonnett}.  To make SN background estimates with high statistical precision, we generate \NGENIa{} SNIa and \NGENCC{} CC \SNe{}; this corresponds to 40 of our \DESSN{} data sets.

\paragraph*{Fakes on nearby galaxies:}
Following K15, we place fake point sources (``fakes") of random magnitudes onto CCD images to determine the efficiency of \diffimg{} in identifying sources on low-redshift galaxies.  While the \snana{} simulation accounts for increased noise from bright galaxies, processing fakes with \diffimg{} includes unmodeled inefficiencies that are not characterized by our simulation. 
As described in K15, there exists a ``surface brightness anomaly" which degrades the \diffimg{} search efficiency for events on nearby galaxies. Pan-STARRS1 has also seen a similar anomaly (see Fig.~6 in \citealt{Rest}).  To determine the effects of this anomaly on our search, we compute and compare the detection efficiencies from the \snana{} and \diffimg+fakes methods for a range of source magnitudes and background surface brightnesses ($\mSB$), where $\mSB$ is the magnitude per square arcsecond measured on the template image. These two methods are used to independently quantify the results of our \KN{} search, which we present in \S~\ref{sec:results}. Since our analysis relies on $i$ and $z$-band detections, the product of the $i$ and $z$-band efficiencies gives the trigger efficiency.  Due to limited resources and a small number of low-redshift galaxies ($\sim100$) in the \SN{} fields, we only test fake foreground magnitudes from 21st to 25th magnitude, and we do not run the difference imaging pipeline on fake \BK{} light curves themselves.

\paragraph*{Not Simulated:}
\citet{Cowperthwaite} (hereafter CB15) identify a number of objects other than \SNe{} that could be \KNe{} contaminants.  These backgrounds are not simulated due to their low expected observation rate and are mostly removed with an $i-z$ color cut (see Fig. 3 in CB15).  CB15 also found background objects with redder $i-z$ colors that could be consistent with the \KN{} color: the Pan-STARRS fast transient PS1-13ess with $i-z = 0.62$ mag \citep{PS1} and Type ``.Ia" \SNe{} \citep{Shen}.   Tables 1 and 2 in CB15 show that the observation rate of these contaminants is about two orders of magnitude lower than the \SN{} Ia observation rate, but still enough to contaminate \KN{} signals. Nevertheless, these objects fail our selection requirements as described in \S~\ref{subsec:notsim}.

\section{Analysis}
\label{sec:analysis}
Our analysis extracts volumetric rest-frame rate limits from the \DESSN{} data sample using estimates of search time-volume and search efficiency.  Following \citet{Dilday} (hereafter D08), the average \KN{} rest-frame volumetric rate we infer is given by
\begin{equation}
R = \NKN / [ \widetilde{\effKN VT}],
\label{eq:rate}
\end{equation}   
where $\widetilde{\effKN VT}$ is the effective time-volume probed by the survey for a volume $V$, observation time $T$, and efficiency as a function of redshift $\effKN(z)$.  $ \widetilde{\effKN VT}$ is computed from D08:
\begin{equation}
\widetilde{\effKN VT} = (\Theta T)\int_{z_{min}}^{z_{max}} dz \epsilon(z)u^2(z)\frac{du}{dz}\frac{1}{(1+z)}
\label{eq:VTeff}
\end{equation}   
Here, $\Theta$ is the solid angle probed by the survey in all fields and $u$ is the FLRW metric comoving distance. The \KN{} efficiency $\epsilon(z)$ is defined as the fraction of simulated \KNe{} at redshift $z$ passing the selection requirements described in \S~\ref{sec:cuts}. Dependence of $\epsilon(z)$ on factors such as cadence and observing conditions are integrated out over the two \DES{} seasons. We do not consider local density perturbations, which could affect volumetric rate estimates at low redshifts. 

\subsection{Selection Requirements}
\label{sec:cuts}
Here we describe the selection requirements (cuts) applied to the data and simulated events.  The cuts were designed to exclude \SNe{} while maximizing \KN{} efficiency. 
\begin{enumerate}
	\item A {\it \KN{} trigger} requires paired $i$- and $z$-band detections, where a detection is described in K15. Both detections pass automated image scanning described in \citet{Goldstein}. We also require that both detections have a signal-to-noise ratio (SNR) greater than 5.  The $i$- and $z$-band detections must occur on the same night for a \KN{} trigger except in the deep fields where adjacent night detections are allowed due to longer deep-field exposure times.  A candidate can only \KN{} trigger once: the first paired $i+z$ detection of a candidate is taken as the \KN{} trigger and subsequent $i+z$ detections of the same candidate are not considered additional triggers.
	\item For each $i$- and $z$-band detection in cut 1, we require zero point (ZP) $>$ \ZPmedian$-0.2$ mag: \ZPmedian{} is the median of the ZP distribution for KN trigger $i$ and $z$ detections, and it is calculated independently for the shallow and deep fields. This cut ensures reasonable atmospheric transparency during the \KN{} trigger observations.
	\item For each $i$- and $z$-band detection in cut 1, we require PSF full width at half maximum $<$ 2.0 arcsec.
	\item The ratio $g$-flux/$z$-flux $<$ 0.15 on the \KN{} trigger night.  If the $i$ and $z$ bands are observed on adjacent nights (in the deep fields), the $g$-band flux on the earliest night of the \KN{} trigger is used.  Considering all nine \BK{} models, the $g$-flux is at most 0.09 times the $z$-flux at time of peak $z$-flux.  The cut is well above the maximum $g$-flux/$z$-flux to accept events with reasonably large Poisson fluctuations.
	\item $r$-flux/$z$-flux $<$ 0.4: Similar to the $g$-flux/$z$-flux cut except using the $r$ band rather than the $g$ band. Considering all nine \BK{} models, the $r$-flux is at most 0.37 times the $z$-flux at time of peak $z$-flux.
	\item There is at least one $i$ or $z$ band observation (regardless of {\small SNR}) 3 to 10 days\footnote{All temporal cuts are made in the observer frame.} after the \KN{} trigger: This ensures sufficient data to examine the light curve evolution.
	\item There is at least one observation (independent of {\small SNR}) 2 to 14 days before the trigger: ensures the ability to identify (and reject) light curves that begin before the trigger.
	\item There is at least one observation (independent of {\small SNR}) 20 to 100 days after the trigger: ensures the ability to identify (and reject) light curves that continue after any \KN{} would have faded.
	\item There is no {\small SNR} $>$ 4 observation 2 to 14 days before the trigger: rejects objects that are bright before the \KN{} trigger.
	\item There is no {\small SNR} $>$ 4 observation more than 20 days after the trigger:  We define the time difference between the last single-band detection and the \KN{} trigger $\Delta t \equiv t_{\rm last} - t_{\rm trigger}$ and require $\Delta t < 20$ days to reject objects with long timescales.  
	\item Veto events matched to a host galaxy with \zphot{} $>$ 0.3: Since we are sensitive to \KNe{} at low redshift ($z < 0.3$), we remove events associated with a high redshift galaxy.  The requirement for matching a source to a host galaxy is a source-host galaxy angular separation $< 2''$ and a directional light radius separation of $d_{LR} < 2$ \citep{Sako,Gupta}.  This matching requirement is more strict than the K15 requirement of $d_{LR} < 4$, and was adjusted to avoid too many false matches to hosts, which lowers the \KN{} efficiency. If a source is matched to a host, it is vetoed if the host has \zphot{} $>$ 0.3. Given the \DESSN{} limiting magnitudes, transients detected with \zphot{} $>$ 0.3 have peak absolute magnitude brighter than $\sim-16.5$ mag in the deep fields and $-17.5$ mag in the shallow. Note that transients with no detected host galaxy will have no \zphot{} and thus this veto will not apply.
	\item shape $<$ 0.  Here we define the `shape' as the rate of change of the $z$-band flux (normalized to the \KN{} trigger $z$-band flux) between the night of the \KN{} trigger and the next $z$-band observation:
	\begin{equation}
	\textrm{shape} = \frac{1}{F_{1}} \frac{F_2-F_{1}}{t_{2}-t_1}, 
	\label{eq:shape}
	\end{equation}
	where $F_{1}$ and $t_1$ are the $z$-band flux and MJD of the \KN{} trigger, respectively. $F_2$ and $t_2$ are the flux and MJD of the next $z$-band observation after the \KN{} trigger, respectively. This shape cut removes events that do not exhibit a declining light curve after the trigger.  
	\item {\it angular separation} (\diz) between the \KN{} trigger $i$ and $z$ observations of the transient $<$ 0.6":  This cut removes contamination from asteroids.  The \diz{} distribution is shown in Fig.~\ref{fig:AngsepNew} for the deep and shallow fields after applying the first 10 cuts. The deep field exposure times for the $i$ and $z$ bands are 1800 and 3600 seconds respectively, long enough that moving asteroids fail the PSF-shape requirement.  The deep field sample is thus dominated by non-moving transients, and the \diz{} distribution is peaked near 0.1", consistent with the astrometric precision.  The shallow field exposure times are much shorter, 200 and 400 seconds for $i$ and $z$ band respectively, and thus slow-moving asteroids are included by the 1" trigger-matching requirement in \diffimg.  Fig.~\ref{fig:AngsepNew} shows that the shallow field \diz{} distribution has two components: 1) a non-moving component at \diz{} $\sim$ 0.1 and 2) a moving component with more events as \diz{} increases.  
	\begin{figure}[t!]
	\centering
	\includegraphics[scale=0.6]{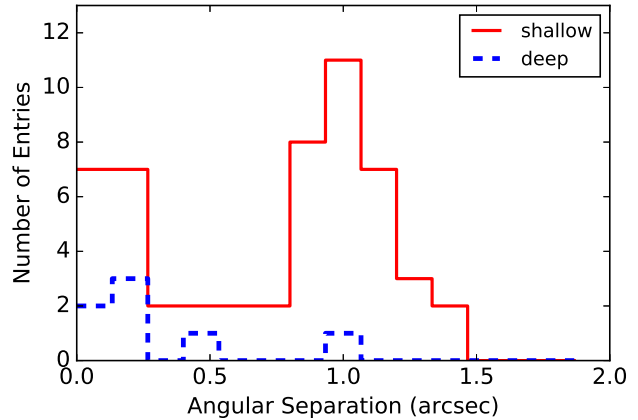}
	\caption{Distributions of the angular separation between \KN{} trigger $i$ and $z$ observations \diz{} for the \DESSN{} data.  The first ten cuts are applied.  A few objects with outlying \diz{} are not shown on the plot.}
	\label{fig:AngsepNew}
\end{figure}
\begin{figure*}[t!]
	\centering
	\includegraphics[scale=0.6]{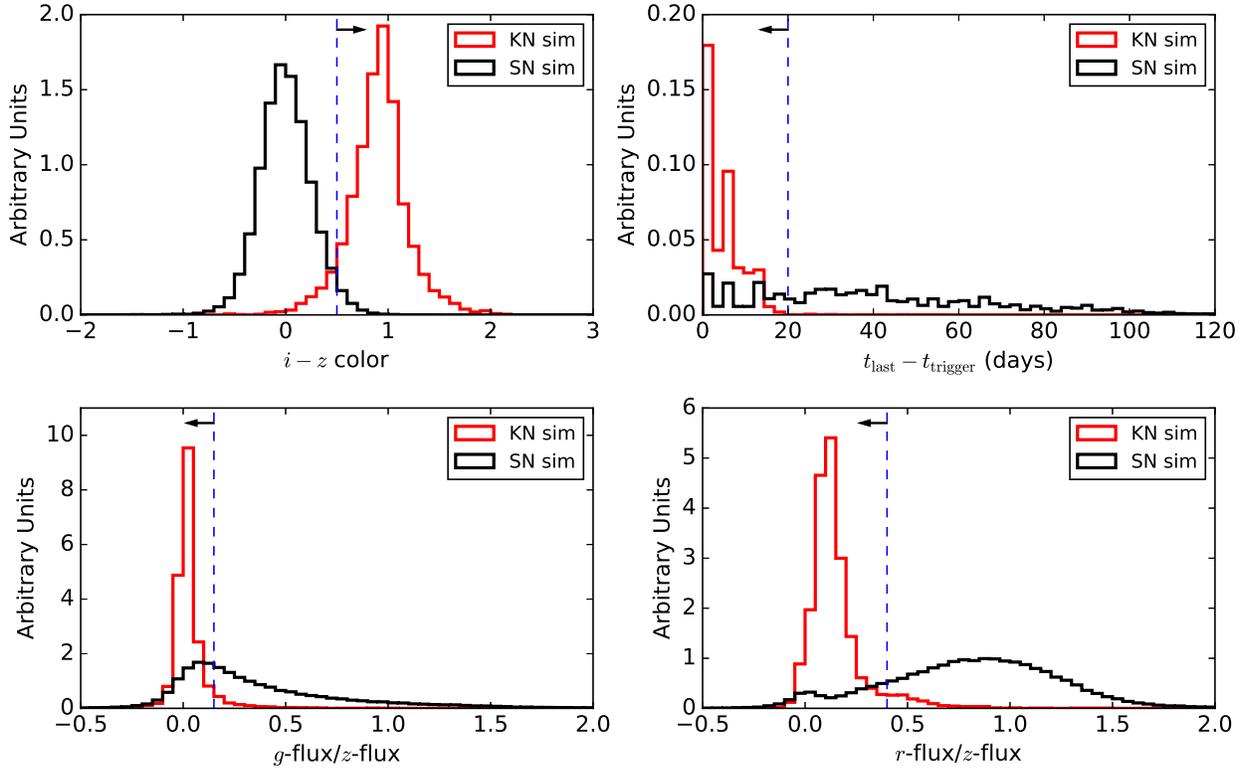}
	\caption{Simulated distributions of KN and SN triggers (cut 1). The dashed vertical lines show the values of the cuts, and the arrows show the selected sample.  {\it Top left panel:} $i-z$ colors.  {\it Top right:} Time between first trigger and last single-band detection. {\it Bottom left:} ratio of $g$-flux to $z$-flux for a trigger.  Negative values are allowed, since negative fluxes can occur due to forced photometry.  {\it Bottom right:} ratio of $r$-flux to $z$-flux for a trigger.}  
	\label{fig:KNSN4panel}
\end{figure*}

\begin{table*}[t!]
\caption{Number of Events and Simulated Efficiencies for each Selection Requirement}
\begin{tabular}{c|M{.6cm} M{.8cm} M{.8cm} M{.8cm} M{.85cm} |M{.6cm} M{.8cm} M{.8cm} M{.8cm} M{.85cm}|}
\cline{2-11}
& \multicolumn{5}{ c| }{Shallow Fields\TBstrut}& \multicolumn{5}{ c| }{Deep Fields\TBstrut} \\ \cline{2-11}
\hline
\multicolumn{1}{|l|}{Cuts} & Data &sim \SN{}\tablenotemark{a}& $\rm eff_{CC}$\tablenotemark{b} & $\rm eff_{Ia}$\tablenotemark{b} & $\rm eff_{\KN{}}$\tablenotemark{b}\tablenotemark{c}& Data & sim SN{}\tablenotemark{a} & $\rm eff_{CC}$\tablenotemark{b} & $\rm eff_{Ia}$\tablenotemark{b} & $\rm eff_{\KN{}}$\tablenotemark{b}\tablenotemark{c} \\  \hline
\multicolumn{1}{|l|}{1. i+z trigger} & 3487 & 1214 & 0.01 & 0.28 & 4.3 & 1236 & 1062 & 0.045 & 0.89 & 3.9 \\
\multicolumn{1}{|l|}{2. ZP $<$ \ZPmedian} & 3300 & 1172 & 0.01 & 0.27 & 4.1 & 1180 & 1010 & 0.043 & 0.85 & 3.7 \\
\multicolumn{1}{|l|}{3. trigger PSF($i$,$z$) $<$ 2.0 "} & 3074 & 1139 & 0.0098 & 0.27 & 3.9 & 1176 & 1004 & 0.042 & 0.84 & 3.6 \\
\multicolumn{1}{|l|}{4. gflux/zflux $<$ 0.15} & 550 & 237 & 0.0019 & 0.057 & 3.7 & 407 & 562 & 0.019 & 0.53 & 3.4 \\
\multicolumn{1}{|l|}{5. rflux/zflux $<$ 0.4} & 209 & 34 & 0.0005 & 0.0058 & 3.6 & 213 & 280 & 0.009 & 0.27 & 3.3 \\
\multicolumn{1}{|l|}{6. observed 3-10 days after trigger} & 195 & 28 & 0.0004 & 0.0045 & 3.1 & 169 & 214 & 0.0069 & 0.21 & 2.9 \\
\multicolumn{1}{|l|}{7. observed 2-14 days before trigger} & 172 & 22 & 0.0003 & 0.0033 & 3 & 152 & 189 & 0.0061 & 0.18 & 2.8 \\
\multicolumn{1}{|l|}{8. observed 20 to 100 days after trigger} & 164 & 20 & 0.0003 & 0.0032 & 2.6 & 148 & 175 & 0.0055 & 0.17 & 2.7 \\
\multicolumn{1}{|l|}{9. no SNR $>$ 4 observed 2 to 14 days before trigger} & 61 & 5.1 & 8$\times 10^{-5}$ & 0.0008 & 2.6 & 24\tablenotemark{d} & 42 & 0.0014 & 0.041 & 2.5 \\
\multicolumn{1}{|l|}{10. no SNR $>$ 4 observed 20 days after trigger} & 53 & 4.0 & 6$\times 10^{-5}$ & 0.0007 & 2.5 & 8 & 11 & 0.0004 & 0.0096 & 2.5 \\
\multicolumn{1}{|l|}{11. veto $z_{phot} >$0.3} & 51 & 1.9 & 3$\times 10^{-5}$ & 0.0004 & 2.3 & 2 & 6.2 & 0.0002 & 0.0058 & 2.2 \\
\multicolumn{1}{|l|}{12. shape $<$ 0.0} & 47 & 1.4 & 2$\times 10^{-5}$ & 0.0002 & 2.2 & 2 & 3.9 & 0.0001 & 0.0035 & 2.2 \\
\multicolumn{1}{|l|}{13. \diz $<$ 0.6 arcsec} & 12 & 1.4 & 2$\times 10^{-5}$  & 0.0002 & 2.2 & 1 & 3.9 & 0.0001 & 0.0035 & 2.2 \\
\multicolumn{1}{|l|}{14. i-z color $>$ 0.5} & 0 & 0.17 & 3$\times 10^{-6}$ & 2$\times 10^{-5}$  & 2.2 & 0 & 0.96 & 3$\times 10^{-5}$  & 0.0009 & 2.1 \\
\hline
\end{tabular}
 \tablenotetext{1}{Since 40 SN data sets are simulated for better statistics, the numbers of simulated SN events are divided by 40 to correspond to one DES-SN data set. The statistical uncertainty in each sim SN line is $\sqrt{N_{\rm SN}/40}$, where $N_{\rm SN}$ is the number of simulated events on a given line.}
 \tablenotetext{2}{$\rm eff_{CC}$, $\rm eff_{Ia}$, and $\rm eff_{\KN{}}$ are the simulated efficiencies of CC SNe, SNIa, and KNe, respectively. They are quoted in percent.}
 \tablenotetext{3}{Based on a random assortment of \BK{} light curves generated randomly in co-moving volume out to redshift 0.15 in shallow fields and 0.2 in deep fields assuming no host galaxy noise. The \KN{} efficiencies shown here are small due to the large redshift range of the simulation.}
 \tablenotetext{4}{The 2.8$\sigma$ difference between the SN simulation prediction and data here is not unexpected: We estimate a $\sim 30$\% chance of there being such a deviation in the table by performing a Monte Carlo simulation of data values in the table given the numbers from the SN simulation.}
\label{table:cuts}
\end{table*}

Further evidence for asteroid contamination is the excess of events in the four shallow fields near the ecliptic where most asteroids orbit.  After applying the first 10 cuts, there are a total of \shallowEcliptic{} events in these four fields, while the four shallow fields away from the ecliptic have a total of \shallowNoEcliptic{} remaining events. In contrast, the two deep fields (one near the ecliptic and one away from the ecliptic) each contain the same number of events (\deepEcliptic{}) after cut 10, showing that the deep fields are less susceptible to asteroid contamination.  
	\item \KN{} trigger $i-z$ color $>$ 0.5 mag: Motivated by the \BK{} predictions that \KNe{} are very red, we require $i-z>0.5$ for the trigger bands.
\end{enumerate}

\begin{table}
\caption{KN efficiency for peak $m_i=18$\tablenotemark{a}}
 \tablenotetext{1}{BK13 $M=0.1M_{\odot}$ models simulated in shallow fields with \snana, where each model is scaled to have peak $i$ band magnitude $m_i=18$.}
\begin{tabular}{l|M{1cm}|M{1cm}|M{1cm}|@{}M{0pt}@{}|}
\cline{2-4}
& \multicolumn{3}{c| }{KN efficiency for\TBstrut} \\[2pt] \cline{2-4}
\hline
\multicolumn{1}{|l|}{Cut} & $\beta=0.1$ & $\beta=0.2$ & $\beta=0.3$ &\\[5pt] \hline
\multicolumn{1}{|l|}{1. i+z trigger} & 0.99 & 0.99 & 0.99 &\\[5pt]
\multicolumn{1}{|l|}{2. ZP $<$ \ZPmedian} & 0.88 & 0.90 & 0.88 &\\[5pt]  
\multicolumn{1}{|l|}{3. trigger PSF($i$,$z$) $<$ 2.0 arcsec} & 0.77 & 0.77 & 0.78 &\\[5pt]  
\multicolumn{1}{|l|}{4. gflux/zflux $<$ 0.15} & 0.64 & 0.72 & 0.78 &\\[5pt]  
\multicolumn{1}{|l|}{5. rflux/zflux $<$ 0.4} & 0.59 & 0.72 & 0.78 &\\ [5pt] 
\multicolumn{1}{|m{4cm}|}{6. observed 3 to 10 days after trigger} & 0.55 & 0.69 & 0.73 &\\[15pt]  
\multicolumn{1}{|m{4cm}|}{7. observed 2 to 14 days before trigger} & 0.53 & 0.66 & 0.71 &\\[15pt]   
\multicolumn{1}{|m{4cm}|}{8. observed 20 to 100 days after trigger} & 0.47 & 0.60 & 0.64 &\\[15pt]  
\multicolumn{1}{|m{4cm}|}{9. no SNR $>$ 4 observed 2 to 14 days before trigger} & 0.45 & 0.58 & 0.63 &\\[15pt] 
\multicolumn{1}{|m{4cm}|}{10. no SNR $>$ 4 observed $>$ 20 days after trigger} & 0.44 & 0.58 & 0.62 &\\[15pt] 
\multicolumn{1}{|m{4cm}|}{11. veto $z_{phot} >$0.3} & 0.41 & 0.53 & 0.55 &\\[5pt] 
\multicolumn{1}{|l|}{12. shape $<$ 0.0} & 0.40 & 0.50 & 0.55 &\\[5pt]  
\multicolumn{1}{|l|}{13. \diz $<$ 0.6 arcsec} & 0.40 & 0.50 & 0.55 &\\[5pt]  
\multicolumn{1}{|l|}{14. i-z color $>$ 0.5 mag} & 0.40 & 0.50 & 0.55 &\\[5pt]  
\hline
\end{tabular}
\label{table:cutsvsmodel}
\end{table}

  \begin{figure*}[t!]
	\centering
	\includegraphics[scale=0.43]{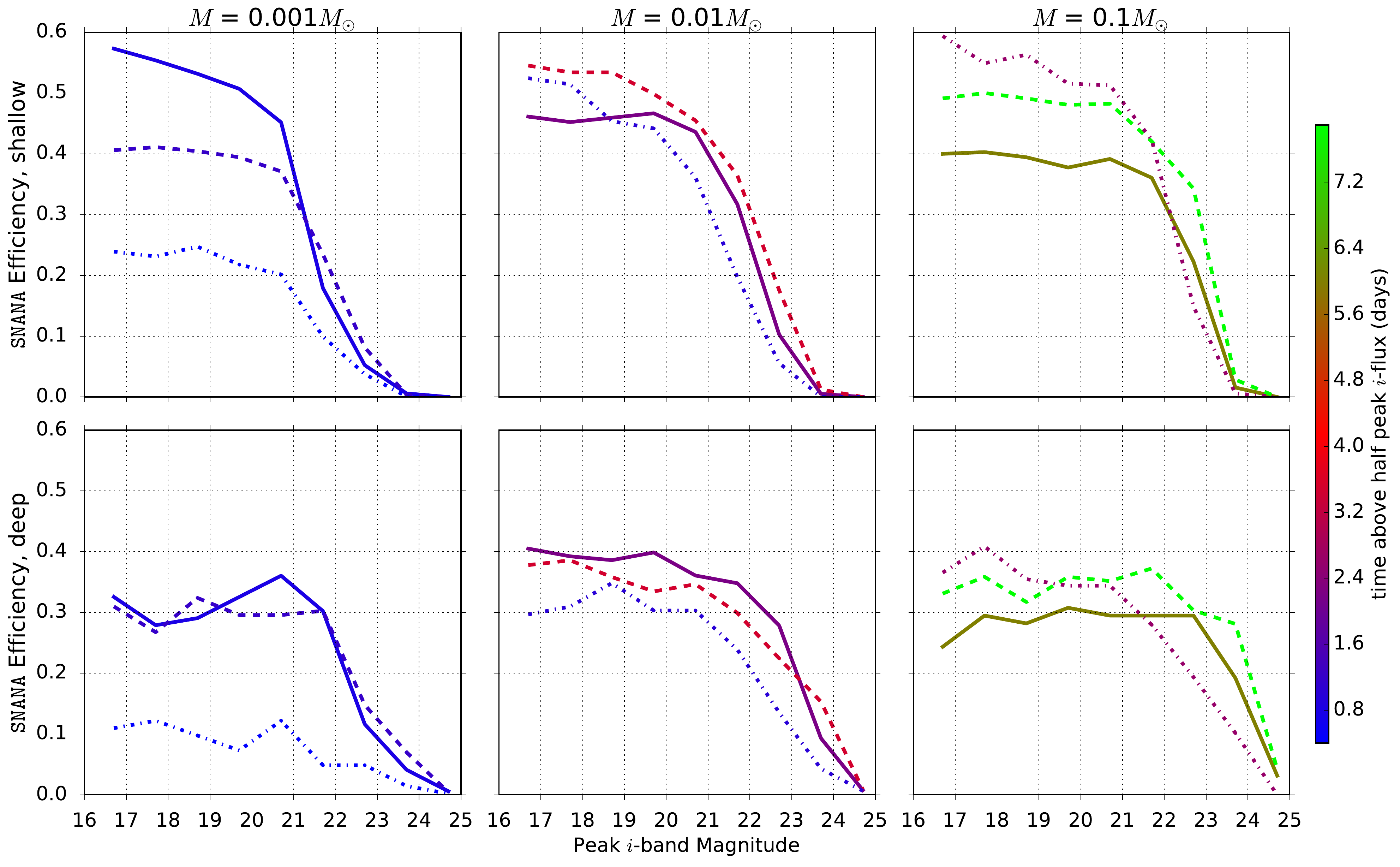}
	\caption{\KN{} search efficiency as a function of peak apparent $i$-band magnitude for the \BK{} models in the shallow and deep fields at fiducial redshift $z=0.02$.  Efficiencies do not account for host galaxy noise.  The solid, dashed, and dashed-dotted lines correspond to \BK{} models with $\beta = 0.1, 0.2, 0.3$, respectively.  The colors of the lines represent each model's $t_{\rm half}$, the time above half maximum flux in the $i$-band.}
	\label{fig:EffVsMag}
\end{figure*}

We exclude one cut at a time from the analysis to determine the effectiveness of each cut.  We find that the four cuts which maximally suppress \SN{} background, in order of effectiveness, are cuts 9, 14, 10, and 11.  Though the $g$- and $r$-band cuts (4 and 5) are not among the most effective cuts for \SNe{}, we still require these color cuts to reject other backgrounds, as described in \S~\ref{subsec:notsim}.  The simulated distributions of the three colors (cuts 4, 5, and 14) and $\Delta t$  (cut 10) are shown in Fig.~\ref{fig:KNSN4panel} for \KNe{} and \SNe{}. The \KN{} distributions shown here are based on a random assortment of the nine \BK{} models placed uniformly in co-moving volume out to $z=0.15$ in the shallow fields and $z=0.2$ in the deep fields.

The number of data and simulated events remaining after each cut is shown in Table~\ref{table:cuts}.  No events from the \DESSN{} sample pass all of our cuts, and the \SN{} simulation predicts \NCUTSSIM{} total background events.  

\subsection{Efficiency for Transients that are Not Simulated}
\label{subsec:notsim}
The cuts are tuned solely on the \SN{} simulations but are effective at removing other backgrounds as well.  As mentioned in \S\ref{sec:simulations}, Type ``.Ia" \SNe{} and Pan-STARRS fast transients like PS1-13ess have $i-z$ colors consistent with \KNe{}. PS1-13ess is bright in the $g$-band relative to the $z$-band \citep{PS1}, so objects like it are rejected with a cut on the $g$ brightness (cut 4).   ``.Ia" models from \citet{Shen} produce spectra that are initially a blue continuum, but redden at late times as the model dims.  Such light curves are removed with cuts 4, 5, and 9, which remove events with bright $r$ or $g$-band flux before or during the \KN{} trigger.  

\subsection{\KN{} efficiency with no host galaxy} 
\label{subsec:nohostgal}
  \begin{figure*}[t!]
	\centering
	\includegraphics[scale=0.4]{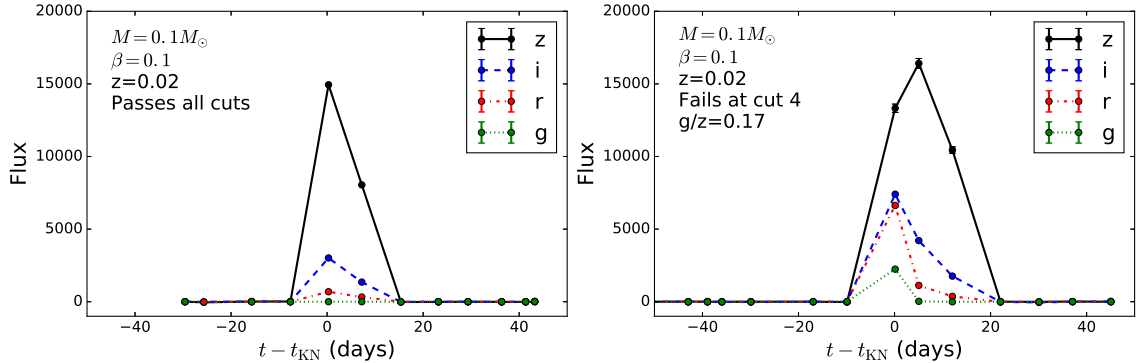}
	\caption{Example {\it griz} $\beta=0.1$, $M=0.1M_{\odot}$ light curves at magnitude 18 that pass the cuts (left) and fail the cuts at cut 4 (right).  The points show the observed fluxes and lines are drawn between them to guide the eye.  Fluxes are defined such that magnitudes are given by $27.5 - 2.5\log_{10}({\rm Flux})$; e.g., mag=22.5 for Flux=100.}
	\label{fig:gzcutlc}
\end{figure*}

To calculate \KN{} rates using Equations \ref{eq:rate} and \ref{eq:VTeff}, the \KN{} efficiency as a function of redshift or apparent magnitude is required.  Fig.~\ref{fig:EffVsMag} shows the efficiency versus peak apparent magnitude for the nine \BK{} models. To generate these curves, we vary the absolute magnitude of \BK{} light curves and simulate them at a fiducial redshift of $z=0.02$ with no host galaxy noise. Here, a fiducial redshift is used so that the dependence of efficiency on observed brightness can be compared between models, while being agnostic to each model's absolute magnitude. Our final rate calculations, however, use analogous efficiency curves and include redshifting based on absolute magnitudes from \BK{}.  At bright apparent magnitudes, the efficiencies do not reach unity due to the \DESSN{} cadence and selection requirements. 

To further illustrate selection effects, Fig.~\ref{fig:EffVsMag} shows the timescale of each \KN{} model, parameterized by the time above half maximum flux in the $i$ band $t_{\rm half}$.  Near the detection limit (mag 23.5 for shallow fields, mag 24.5 for deep), \KNe{} with longer $t_{\rm half}$ are more detectable, because there are more chances to make a detection than for short-$t_{\rm half}$ \KNe{}.  However, this relation between efficiency and $t_{\rm half}$ does not hold at magnitudes much brighter than the detection limit.  For a bright \KN{}, there are multiple chances to detect the light curve before it falls below the detection threshold, so the selection requirements rather than $t_{\rm half}$ drive the efficiency for each model.  

For comparison between models, Table~\ref{table:cutsvsmodel} shows the efficiency of each $M=0.1M_{\odot}$ model at magnitude 18 ($\sim 5$ mag brighter than threshold) and $z=0.02$ in the shallow fields with each cut.  The cut on the ratio of g-flux to z-flux (cut 4) significantly reduces the $\beta=0.1$, $M=0.1M_{\odot}$ efficiency, showing how the interplay between color and cuts will affect the efficiency of bright sources. Although cut 4 diminishes the $\beta=0.1$, $M=0.1M_{\odot}$ efficiency, the cut value of $g/z=0.15$ was chosen to maximize the signal over all nine \BK{} models while removing the background \SNe{}.  Example 18th magnitude $\beta=0.1$, $M=0.1M_{\odot}$ light curves are shown in Fig.~\ref{fig:gzcutlc}.  The light curve in the left panel of Fig.~\ref{fig:gzcutlc} passes the cuts, while that in the right panel fails at cut 4 because of its high g-flux on the detection night.  The discrepancy in measured g-flux between these two light curves is due to the timing of the observations with respect to the light curve start times, exemplifying the effect of the cadence on the search efficiency.

\subsection{\KN{} efficiency with underlying host galaxy}
\label{subsec:hostgal}
  \begin{figure*}[t!]
	\centering
	\includegraphics[scale=0.5]{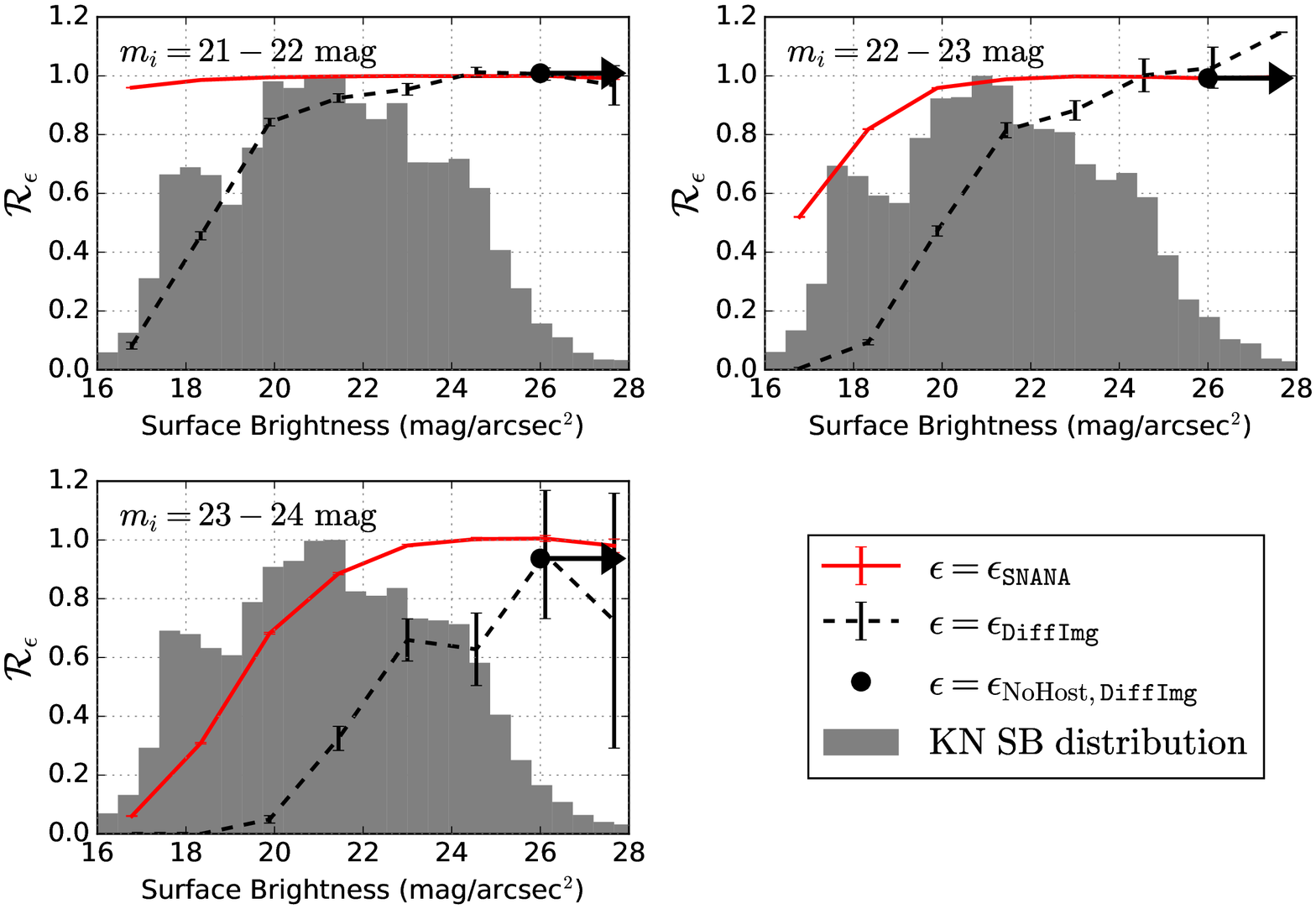}
	\caption{$\effratio$ in the \DESSN{} shallow fields vs. $\mSB$ for \KN{} $i$-band magnitudes (a) $m_i =$ 21-22, (b) 22-23, and (c) 23-24. The solid red line is based on the \snana{} simulation, and the dashed black line is from fake point sources processed by \diffimg{}.  The gray regions show the $i$-band surface brightness distribution in the shallow-field, calculated in \snana{} from the \DES{}+2MASS catalogue Sersic profiles. The black dot and arrow shows $\effratio$ for \diffimg{} fakes with $\mSB>26$, i.e. with no host galaxy. The error bars show 1$\sigma$ uncertainties.}
	\label{fig:DiffimgVsSNANA}
\end{figure*}

Here we investigate two anomalous effects from difference imaging on bright galaxies that result in efficiency losses: 1) excess missed detections of point sources, and 2) excess flux scatter which affects analysis selection requirement efficiency.  For the first effect, poor subtractions can result in \KN{} events that are not detected or that result in detections which look like mis-subtractions and thus fail the automated scanning. To investigate the detection efficiency losses, we use fakes to measure single-band, single-epoch {\it detection} efficiencies for \diffimg{} ($\effdiffimg$\footnote{$\epsilon$ includes the host galaxy noise unless a ``noHost" subscript is included}) as a function of background surface brightness ($\mSB$). The single-epoch efficiency in the \snana{} simulation ($\effsnana$) accounts for host galaxy Poisson noise as well as excess flux scatter from the SB anomaly, but does not account for detection losses.  We therefore use the \diffimg{} detection efficiencies, measured with fake sources on galaxies, to correct $\effsnana$. 

The second image-subtraction effect is from detected \KNe{} with excess flux scatter.  We characterize the impact on the analysis efficiency by including this excess scatter in the simulation (see Figs. 9-10 in K15). In this analysis, we do not consider the effect of correlations between measured fluxes in different bands and correlations between fluxes and detections.  Additionally, our analysis does not account for the effects of image subtraction artifacts on \diz.

To characterize the detection efficiencies with and without a host galaxy, and to compare \snana{} and \diffimg{}, we define an efficiency ratio,
\begin{equation}
\effratio \equiv \epsilon/\effnohost
\end{equation}
where $\epsilon$ is the detection efficiency of either \snana{} or \diffimg{} and $\effnohost$ is the \snana{} efficiency with no host galaxy.  Since $\effnohost$ is the highest possible efficiency, we expect $\effratio \leq 1$, except for statistical fluctuations. Fig.~\ref{fig:DiffimgVsSNANA} shows the ratio $\effratio$ for the $i$-band shallow fields, where $\epsilon$ is 1) $\effsnana$, 2) $\effnohostd$, and 3) $\effdiffimg$. To compute our \KN{} search sensitivity, we have similar information for the $z$-band and the deep fields. 

The simulation of $\effsnana$ shown in Fig.~\ref{fig:DiffimgVsSNANA} demonstrates that host Poisson noise has almost no effect on the detection efficiency for source magnitude $m_i = 21-22$ mag. However, near the detection limit ($m_i=23-24$ mag), $\effsnana$ falls to half the no-host efficiency when the surface brightness reaches $\mSB\sim19$ mag/asec$^2$, which is comparable to sky noise of $19.6$ mag/asec$^2$. The efficiency loss from host Poisson noise can not be mitigated, and represents the upper limit on the efficiency from \diffimg.

\begin{figure}[h!]
	\centering
	\includegraphics[scale=.17]{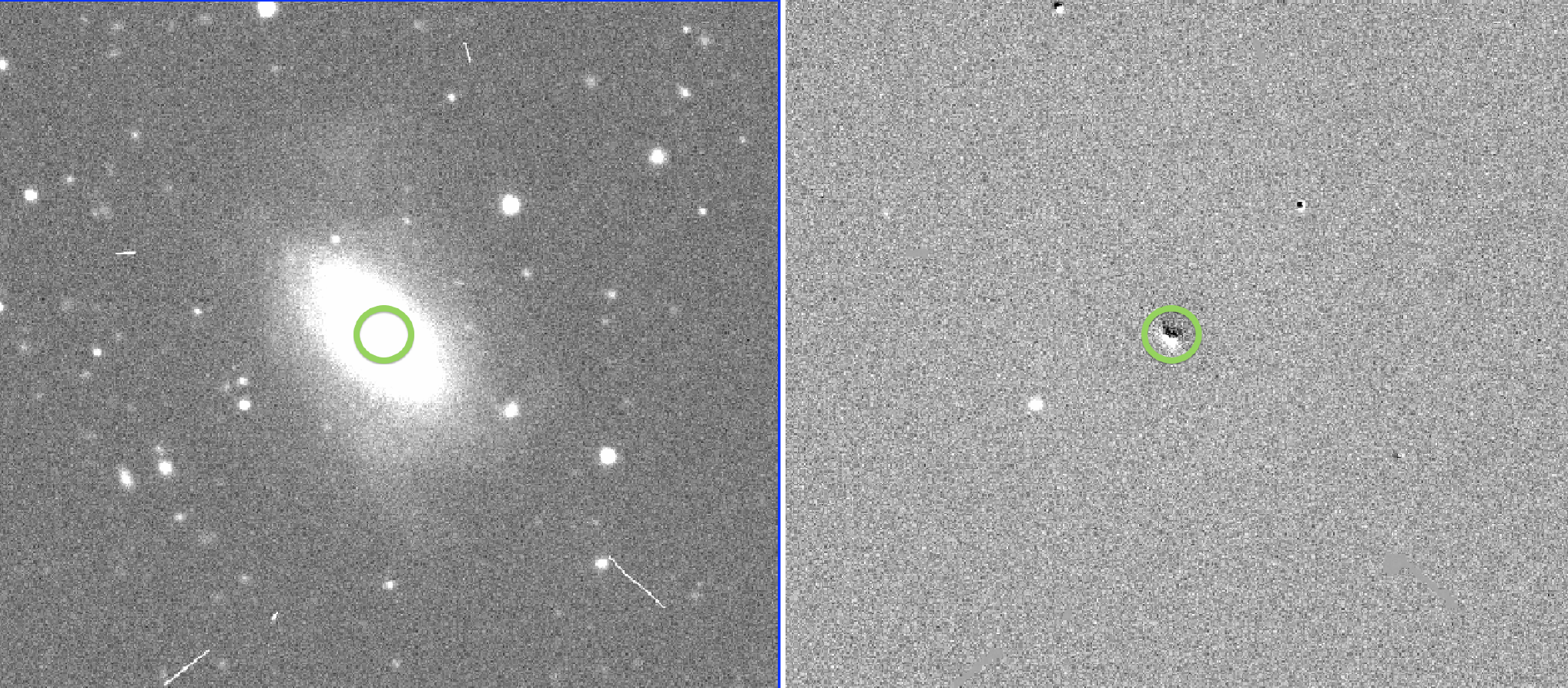}
	\caption{Search (left) and difference (right) images for an undetected fake with source magnitude $m_i=21.2$ mag and $\mSB=17.6$ mag/asec$^2$.  The fake is shown at the center of the green circle. Each image is 2.3'$\times$2.3'. } 
	\label{fig:fakemagimage}
\end{figure}

Fig.~\ref{fig:DiffimgVsSNANA} also shows that the \diffimg{} efficiency with no host galaxy noise $\effnohostd$ agrees with the no-host efficiencies predicted using \snana{} ($\effnohost$).  For $\mSB\gtrsim26$, $\effsnana$ and $\effdiffimg$ converge to $\effnohost$, demonstrating that very faint backgrounds do not degrade efficiency. For brighter backgrounds, the \diffimg{} efficiency ($\effdiffimg$) deviates from the \snana{} efficiency ($\effsnana$) that simulates Poisson noise. At a given surface brightness, $\effdiffimg$ is lower than $\effsnana$, because of the SB anomaly described in K15\footnote{Note that K15 characterized excess scatter in the measured flux, not degraded efficiency: since there are many opportunities to detect the light curve of a bright SNIa, SNIa detection efficiency is not affected by the SB anomaly.}.    For large background brightnesses, the SB anomaly has a significant effect on the efficiency. For fakes with 21 mag $< m_i <$ 22 mag and 17 mag $< \mSB <$ 18 mag/asec$^2$, $\sim60\%$ are detected on the subtracted image, but only $2\%$ of these pass automated image scanning because of poor quality subtractions.  An example search and difference image for an undetected fake on a bright galaxy is shown in Fig.~\ref{fig:fakemagimage}.  This particular fake has an $i$-band magnitude of 21.2 mag and $\mSB=17.6$ mag/asec$^2$. The subtracted image shows a dipole structure and fails the automated scanning.

To estimate the total search sensitivity degradation from host galaxy noise, the efficiency is weighted by the \KN{} distribution of background surface brightness $\mSB$\footnote{Uncertainties on efficiency are shown in Fig.~\ref{fig:DiffimgVsSNANA} for illustrative purposes, but are not propagated through the analysis since host galaxy model uncertainties dominate.}. This distribution has not been predicted, so here we generate \KNe{} at surface brightnesses following the Sersic profiles of the ``\DES{}+2MASS" and SV galaxy catalogues. This procedure assumes that the \KN{} rate density follows the background surface brightness. The \DES{}+2MASS catalogue includes galaxies from the 2MPZ catalogue within 200 Mpc \citep{Bilicki}.  We use the following information from the galaxy catalogues: sky coordinates, redshift, $i$- and $z$-band magnitude, absolute magnitude, and shape profile.  The gray-shaded region in Fig.~\ref{fig:DiffimgVsSNANA} shows the $i$-band surface brightness distribution for \KNe{} derived from the \DES{}+2MASS+SV galaxy catalogue Sersic fits assuming the \KN{} volumetric rate follows galaxy profiles.  While the $\mSB$ distribution depends on the choice of galaxy catalog and model assumptions, the efficiency curves are much less sensitive to these choices. Additionally, we have not investigated potential \diz{} artifacts from image subtractions.

We next compute the average $\effdiffimg / \effsnana$ (ie. excess detection efficiency loss) as a function of \KN{} magnitude weighted by the $\mSB$ distribution.  For $\mSB$ below and above where $\effdiffimg$ is known, the efficiency ratio is set to 0 and {\tt max}($\effdiffimg / \effsnana$), respectively. We then use \snana{} to simulate analogous efficiency curves to those in Fig.~\ref{fig:EffVsMag}, this time including host galaxy Poisson noise and SB-anomaly excess flux scatter.  These new efficiency curves are then down-weighted by the excess detection efficiency loss to give the overall \KN{} efficiency as a function of \KN{} magnitude.  Since the brightest fakes in the \diffimg{} test are $m_i = 21$ mag, we linearly extrapolate $\effdiffimg / \effsnana$ to mag 16 where $\effdiffimg / \effsnana$ is assumed to be $1$.  The recomputed efficiencies which account for the SB anomaly are used to calculate \KN{} volumetric rates in \S~\ref{sec:results}.

\section{Results}
\label{sec:results}

\subsection{Event selection and contamination}
\begin{figure}[t!]
	\centering
	\includegraphics[scale=0.5]{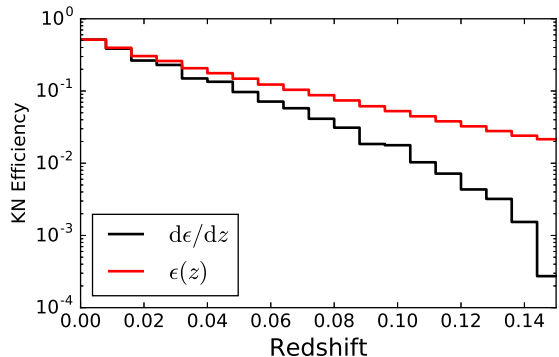}
	\caption{Efficiency per redshift bin ${\rm d}\epsilon / {\rm d}z$ and cumulative efficiency $\epsilon(z)$ in the shallow fields.  } 
	\label{effvsz}
\end{figure}

The number of events passing each cut for the data, simulated \KNe{}, and simulated Ia and CC \SNe{} are shown in Table~\ref{table:cuts}. For the simulations, efficiencies are shown as well. The \KN{} efficiencies in Table~\ref{table:cuts} are based on a different simulation than the 18th magnitude \KN{} efficiencies in Table~\ref{table:cutsvsmodel}. Table~\ref{table:cuts} shows the efficiencies for \BK{} models simulated randomly in co-moving volume out to redshift 0.15 in the shallow fields and 0.2 in the deep fields with no host galaxies. In comparison to Table~\ref{table:cutsvsmodel}, the \KN{} efficiency here is much lower (2\%) because the simulated redshift range goes beyond the detectable distance of most \BK{} models. Fig.~\ref{effvsz} shows the efficiency per redshift bin $({\rm d}\epsilon / {\rm d}z)$ and cumulative efficiency $(\epsilon)$ as a function of redshift for this \KN{} simulation in the shallow fields. The cumulative efficiency falls off as the redshift range is increased, showing that the redshift range and \KN{} absolute magnitude are driving the overall efficiency shown in Table~\ref{table:cuts}.   At redshifts near 0, the efficiency plateaus at $\sim0.35$ rather than 1 because 30\% of the low-redshift \KNe{} do not meet the \KN{} trigger requirement, and there is added loss from the other selection requirements. At $z=0.15$ the cumulative efficiency in Fig.~\ref{effvsz} matches the \KN{} efficiency shown in the shallow column of Table~\ref{table:cuts}.

  \begin{figure*}[t!]
	\centering
	\includegraphics[scale=0.5]{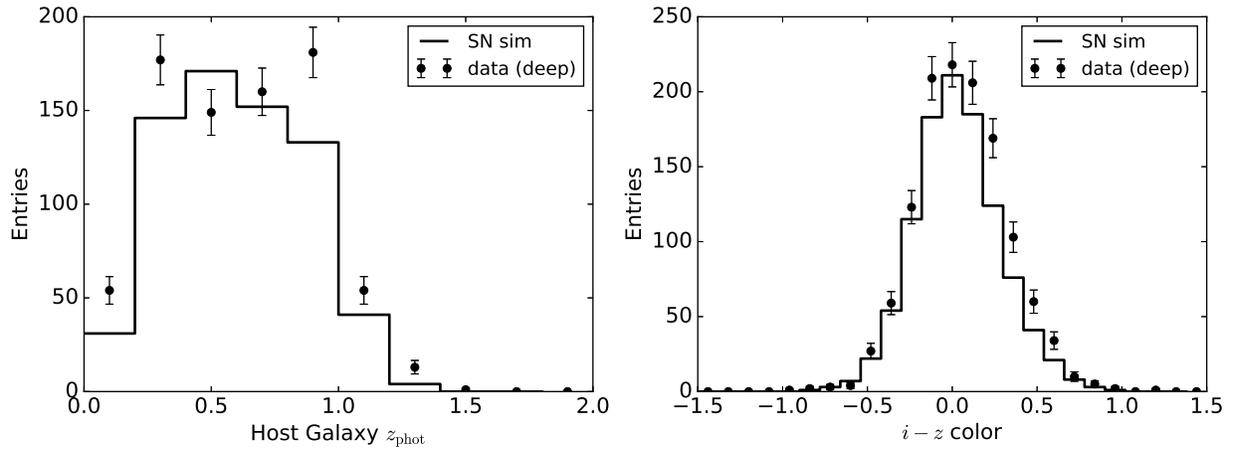}
	\caption{Left: for objects matched to a host galaxy, the distribution of deep field \KN{} trigger \zphot{} is shown for the data (solid circles) and the \SN{} simulation (histogram). Right: Distribution
   of \KN{} trigger i-z color. The simulation has been re-scaled by 1/\SNSIMSCALE{} to correspond to two \DES{} seasons.}
	\label{fig:zphotcolors}
\end{figure*}

During the first two years of \DESSN{}, \shallowTrigs{} and \deepTrigs{} \KN{} triggers were identified in the shallow and deep fields, respectively. As mentioned earlier, asteroids contaminate the shallow fields, so the \SN{} simulations under-predict the number of data triggers. The deep-field triggers however are well described by the \SN{} simulations.   Fig.~\ref{fig:zphotcolors} shows the distributions of host galaxy \zphot{} and $i-z$ color for the \SN{} simulations (normalized to 2 seasons) and the data in the deep fields.  The simulated distributions agree well with those from the data and show that the \SN{} simulations predict the non-asteroid background levels.  After all cuts, no events remain, which is consistent with our simulation prediction of \NCUTSSIM\ \SN{} events: \NPASSSND{} in the deep fields and \NPASSSNS{} in the shallow fields. There is a 33\% chance of finding 0 events in the sample given the \SN{} prediction.

\begin{figure*}[htbp!]
	\centering
	\includegraphics[scale=0.5]{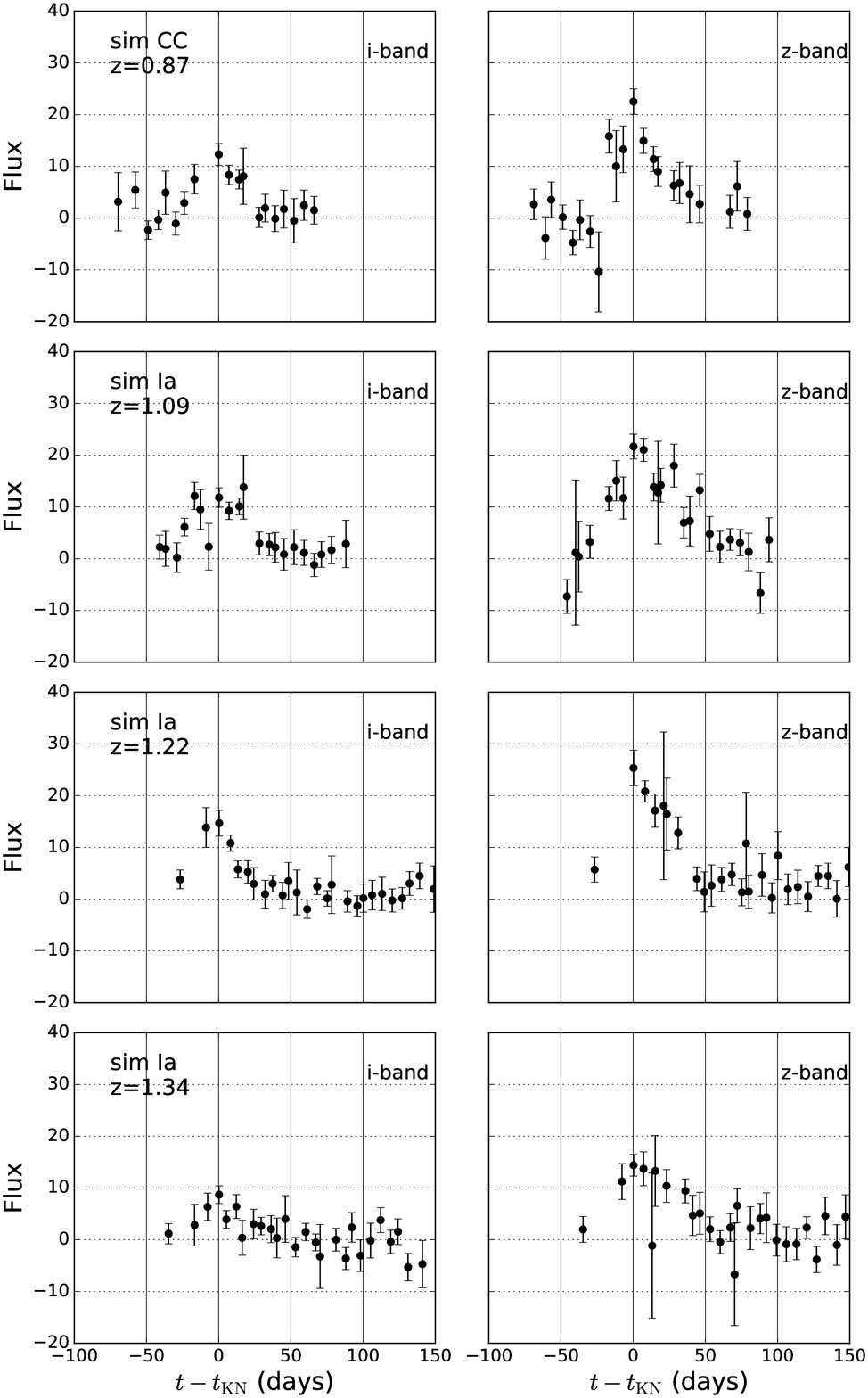}
	\caption{Example $i$ and $z$-band light curves for simulated CC and Ia which passed all cuts.  Fluxes are defined such that magnitudes are given by $27.5 - 2.5\log_{10}({\rm Flux})$; e.g., mag=25 for Flux=10.} 
	\label{passingLCs}
\end{figure*}

\begin{figure*}[t!]
	\centering
	\includegraphics[scale=0.6]{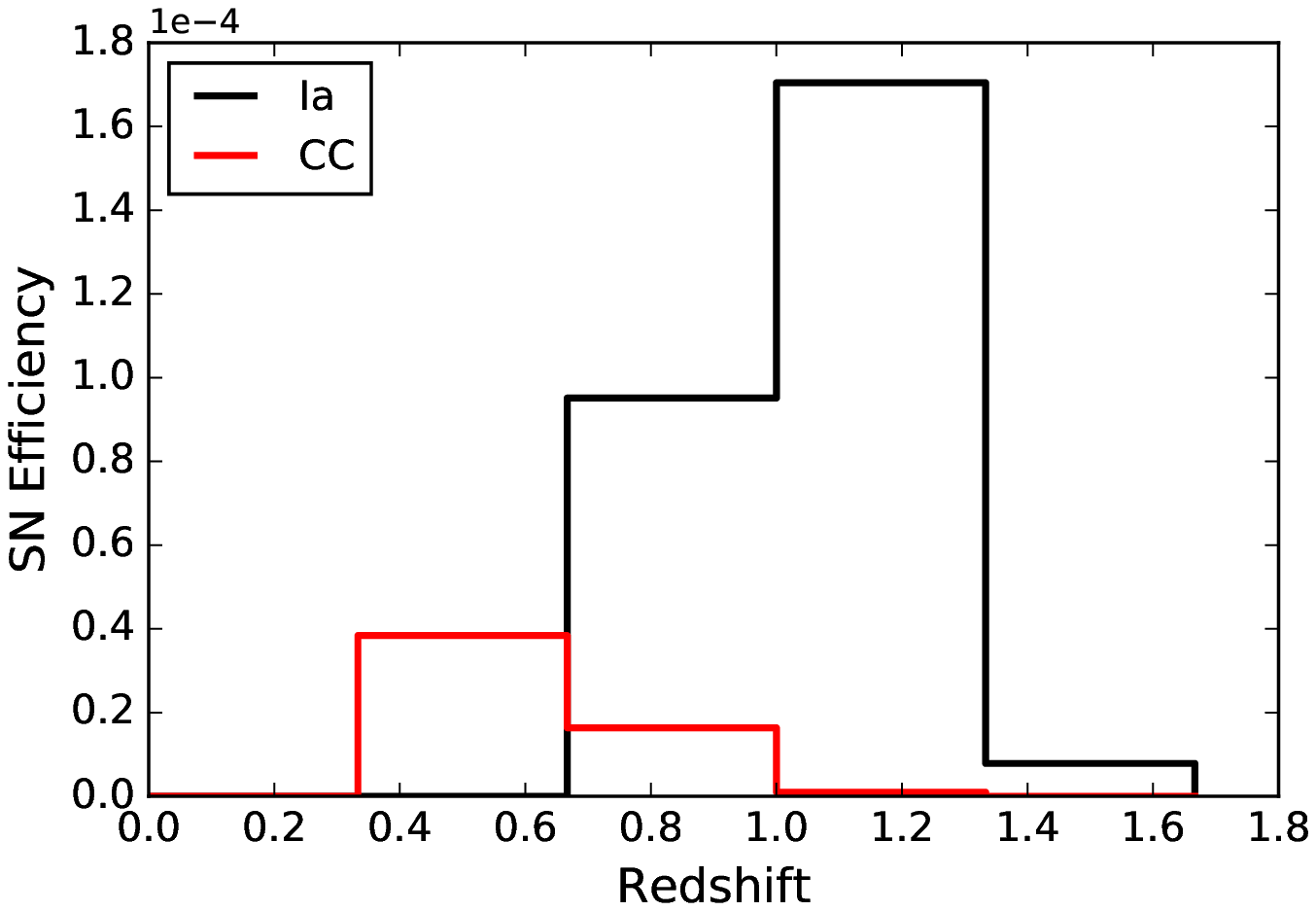}
	\caption{Simulated efficiency of Ia and CC SNe as a function of redshift.  Note that the y-axis is scaled by $10^{-4}$.}  
	\label{fig:SNredshifts}
\end{figure*}

While no events pass our cuts in this search, \SNe{} could contaminate future \KN{} searches (e.g. \citealt{CowperthwaiteFollow}). Fig.~\ref{passingLCs} shows example light curves for simulated CC and Ia which pass all the cuts. About 40\% of the simulated \SN{} background are CC, suggesting that Ia and CC \SNe{} contribute similarly to the \KN{}-search background.  The Ia and CC contaminants have mean redshifts in the simulation of \IaRedshift{} and \CCRedshift{}, respectively, but do not have measured $z_{\rm phot}$ values, and thus can not be vetoed by cut 11. Fig.~\ref{fig:SNredshifts} shows the Ia and CC efficiency as a function of simulated redshift.  At low redshift, the \SNe{} have low efficiency because they exhibit blue light curves and their light curves pass the detection threshold well past 20 days and thus fail the veto (cut 10). However, sufficiently redshifted \SNe{} can take on the red colors of \KN{} light curves and pass cuts 4, 5, and 14.  These simulated contaminants illustrate that high-redshift \SNe{} with no photometric redshift are the primary background for \KN{} searches.


\Needspace{7\baselineskip}
\subsection{Rate limits} \label{ratelims}
\begin{figure*}[p!]
	\centering
	\includegraphics[scale=0.5]{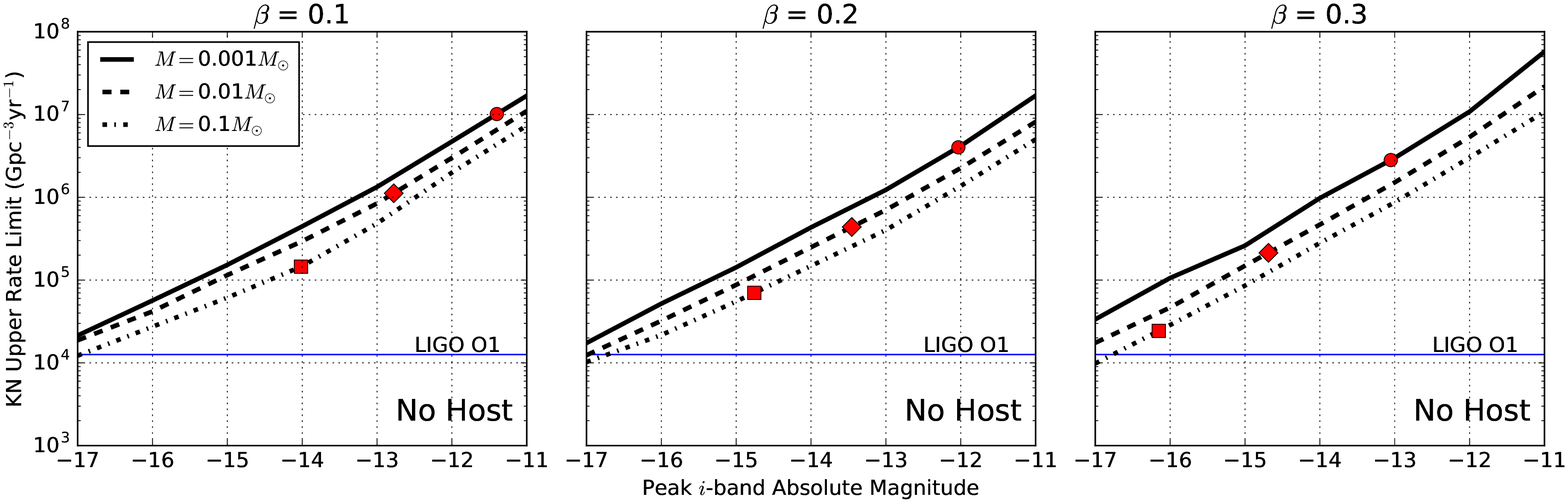}
	\caption{The 90\% upper rate limit for the nine \BK{} models using efficiencies calculated with \snana.  Each \BK{} model is offset in absolute magnitude and the search efficiency is determined to calculate the rate.  The red points show the rate limits for each model with no magnitude offset.  The blue line shows the upper limit set in Advanced \LIGO{} O1 \citep{LIGO}.}
	\label{fig:snanarates}
\end{figure*}

Here we calculate 90$\%$ upper rate limits using Equations~\ref{eq:rate} and ~\ref{eq:VTeff}, efficiencies from \S~\ref{sec:analysis}, and two different assumptions: no host galaxy noise and host galaxy noise including the detection and selection-requirement efficiency loss from the SB anomaly. For these rate calculations, efficiencies differ from those presented in \S~\ref{sec:analysis}, as redshift effects are included rather than simulating at the fixed redshift $z=0.02$.  The first assumption with no host galaxy corresponds to each \BNS{} system being ejected from its host galaxy or occurring in a faint galaxy. For this assumption, we use $\effnohost$ (Fig.~\ref{fig:EffVsMag}) to calculate the rate limits.   The red points in Fig.~\ref{fig:snanarates} show the upper rate limits for the nine \BK{} models assuming all \BNS{} systems are ``kicked" out of galaxies. For the brightest model, the upper rate limit is \minKNratelim{}, while for the dimmest it is \maxKNratelim{}.  We also compute rate limits as a function of absolute magnitude for \BK{} models, which are shown as black lines in Fig.~\ref{fig:snanarates}.

\begin{figure*}[p!]
	\centering
	\includegraphics[scale=0.5]{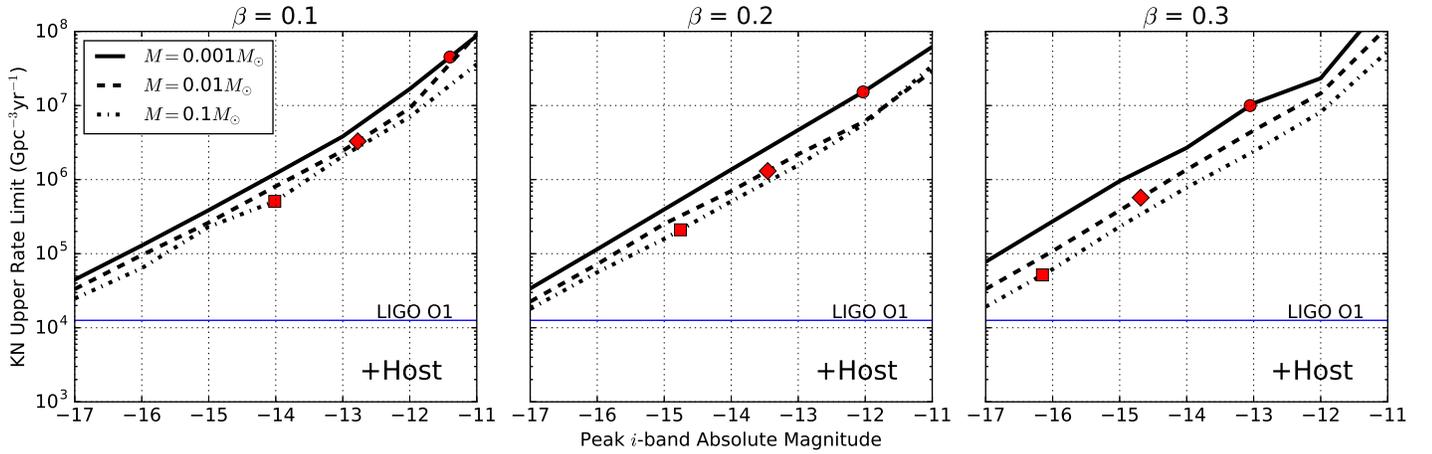}
	\caption{The same as Fig.~\ref{fig:snanarates}, except using \diffimg{} efficiencies accounting for the SB anomaly.}
	\label{fig:diffimgrates}
\end{figure*}

Next we consider the second assumption where all \BNS{} systems remain in the galaxy and are distributed based on background SB, whose distribution is that described in \S~\ref{subsec:hostgal}.  We use $\effdiffimg(z)$ rather than $\effnohost(z)$ for detections and include the excess flux scatter effects in the selection-requirement efficiency.  In this case, the limits are \minKNratelimd{} and \maxKNratelimd{} for the brightest and dimmest \BK{} models, respectively.  Fig.~\ref{fig:diffimgrates} is analogous to Fig.~\ref{fig:snanarates}, but shows the rate limits for \KNe{} in galaxies and includes detection and selection-requirement efficiency loss from Poisson noise and the SB anomaly.  The rate limits on all nine models with and without host galaxy noise are given in Table~\ref{table:ratelimits}. The \KN{} rate upper limits with host galaxy noise are $\sim 3$ times higher than the upper limits for \KNe{} with no detectable host.  The SB anomaly is largely responsible for the decreased sensitivity: the rates calculated with just host Poisson noise are only 10\% higher than with no host galaxy, showing the SB anomaly has a significant impact on the \KN{} search. \\

\begin{table}[htbp!]
\centering
\caption{The 90\% upper rate limits for the BK13 models}
\begin{tabular}{|c|c|c|c|c|}
\hline
\multirow{2}{*}{$M/M_{\odot}$} & \multirow{2}{*}{$\beta$} & \multirow{2}{*}{peak $M_i$} & \multicolumn{2}{|c|}{upper rate limit (Gpc$^{-3}$yr$^{-1}$)} \\ \cline{4-5}
& & & no host & + host \\ \hline
0.001 & 0.1 & $-11.4$ & 1.0 $\times 10^{7}$ & 4.6 $\times 10^{7}$  \\
0.001 & 0.2 & $-12.0$ &  4.0$\times 10^{6}$ & 1.5 $\times 10^{7}$ \\
0.001 & 0.3 & $-13.1$ & 2.8$\times 10^{6}$ & 1.0 $\times 10^{7}$ \\ \hline
0.01 & 0.1 & $-12.8$ & 1.1$\times 10^{6}$ & 3.3 $\times 10^{6}$ \\
0.01 & 0.2 & $-13.5$ & 4.4$\times 10^{5}$ & 1.3 $\times 10^{6}$ \\
0.01 & 0.3 & $-14.7$ & 2.1$\times 10^{5}$ & 5.7 $\times 10^{5}$ \\ \hline
0.1 & 0.1 & $-14.0$ & 1.5$\times 10^{5}$ & 5.1 $\times 10^{5}$ \\
0.1 & 0.2 & $-14.8$ & 7.0$\times 10^{4}$ & 2.1 $\times 10^{5}$ \\
0.1 & 0.3 & $-16.2$ & 2.4$\times 10^{4}$ & 5.2 $\times 10^{4}$ \\ 
\hline
\end{tabular}
\label{table:ratelimits}
\end{table}

\clearpage

\section{Discussion}
\label{sec:discussion}

The upper limits on the \KN{} rate set in our analysis are a few orders of magnitude above other estimates.  The \LIGO{} Collaboration set an upper limit on the \BNS{} merger rate of \ligorate{} \citep{LIGO}, and \citet{FongGRB} estimate a true SGRB rate of \grbrateFongFull{}.  Although the limits presented in this work are not the most stringent, our analysis is independent of and consistent with other experiments and it directly sets limits on \KN-like optical emission. 

A key finding in this work is that the SB subtraction anomaly and \SN{} background signals are major limitations in using DECam to search for \KNe{}. Based on our results, the efficiency loss from underlying host galaxies reduces the \KN{} search sensitivity by a factor of 3 relative to \KNe{} with an undetected host. Further study is also required to determine the sensitivity reduction within the \LIGO{} 80-170 Mpc range \citep{ProspectsLIGO}.  A small part of this reduced sensitivity is from Poisson noise, and cannot be reduced. The major source of degradation, however, is from the difference imaging software. This software was developed and optimized several years ago to discover high-redshift \SNe{}, with little motivation to find faint sources on bright galaxies. With new motivation to find \EM{} counterparts to \LIGO{}-triggered \GW{} events, the software optimizations need to be revisited with the goal of improving the \KN{} efficiency. 

The cuts needed to adequately remove \SNe{} and other backgrounds remove about half of the \KN{} triggers, as shown in Tables \ref{table:cuts} and \ref{table:cutsvsmodel}. Our analysis reduced the predicted \SN{} background to \NCUTSSIM{} events in our sample, but this background will increase with higher search sensitivity or with looser cuts.  While the MJD range of \SN{} explosions will be reduced in follow-up to \LIGO{} triggers, analysis cuts would be relaxed for two reasons. First, to maximize area coverage, only $i$ and $z$ band were used in  \citet{SoaresSantos} (no $g,r$), which removes veto cuts 4 and 5 from \S~\ref{sec:cuts}. Second, there is no pre-trigger veto unless there are serendipitous observations such as in \citet{CowperthwaiteFollow}. They detected a red, rapidly declining source in the follow-up observations of GW151216 \citep{GW151226} with DEcam, but the source was vetoed since the Pan-STARRS Survey for Transients had identified it 94 days before the \GW{} event. These relaxed cuts are likely to increase backgrounds.  

Our analysis also establishes the need for robust asteroid rejection.  The $\sim5$-minute separation between $i$ and $z$ exposures in our shallow-field sample (75 minutes for deep fields) is adequate for removing asteroids.  To maximize the area of future \GW{} follow-up observations, shorter $i$ and $z$ exposure times may be used, and therefore consecutive $i+z$ exposures would be separated by less than the 5 minutes in our \DESSN{} sample. To reduce asteroid contamination, a minimum time between $i$ and $z$ exposures should be considered, such as observing several $i$-band pointings before repeating in the $z$-band. The additional telescope slews may cost more overhead
 than the extra filter changes, and thus EM programs would benefit from a more rigorous analysis of observation strategies.

For future surveys searching for \KNe{}, our analysis has a few important implications. One implication is an estimate of the \KN{} rate limits that could be set with other surveys.  The Large Synoptic Survey Telescope ({\small LSST}), for example, will image $\sim1000$ times more sky area than \DESSN{}, suggesting that {\small LSST} could further constrain the rate limits presented here if the {\small LSST} cadence is comparable to that of \DESSN{} \citep{LSST}.  Our analysis also shows that multi-band observations will be essential for robust background rejection.  Lastly, further optimizations to the survey cadence and temporal spacing of exposures could be made to maximize the gains from a \KN{} search. In particular, a faster cadence would maximize the chance for detection of short-lived \KNe, and a minimum time between the $i$ and $z$ exposures is needed to limit asteroid contamination.

\section{Conclusion}
\label{sec:conclusion}
We have presented a search for \KNe{} in the Dark Energy Survey supernova fields.  Simulations of kilonovae and supernovae were performed to tune cuts, determine search efficiency, and assess background levels.  In our analysis of the first two \DESSN{} seasons, we find no \KN{} candidates and set the first untriggered optical search limits on the rate of \BK{}-like \KN{} events. The most serious issue for our search is a factor of $\sim3$ loss in sensitivity due to difference-imaging for  faint sources on bright galaxies.  For the brightest \KNe{} considered in this analysis, our limits are comparable to the limits set by the \LIGO{} collaboration from gravitational wave observations \citep{LIGO}.  During the course of our analysis, \citet{Barnes2016} updated the \BK{} \KN{} light curve models to account for the efficiency with which radioactive decay products thermalize the ejecta. The light curves for these updated models ({\small BK}16) are still characteristically red, but are fainter compared to \BK{}, especially at late times. At peak, the {\small BK}16 fiducial model bolometric luminosity is roughly half of that predicted without accounting for thermalization efficiency, which sets {\small BK}16 rate limits from our search $\gtrsim 3$ times higher than the \BK{} limits.  Since the {\small BK}16 models dim faster than those of \BK{}, a {\small BK}16 search in our data would suffer additional sensitivity loss from missed detections.  Like \BK{}, the {\small BK}16 models are brightest in the infrared, so the basic methods and cuts presented herein are still applicable to a {\small BK}16 model search, though re-optimization of the cut values could be performed. This work sets the stage for further \KN{} searches with DECam and other large-field-of-view telescopes such as Pan-STARRS1 or Large Synoptic Survey Telescope.  

\section{Acknowledgments}
We would like to thank Dan Kasen for providing us with the \BK{} light curves. We also gratefully acknowledge support from the Kavli Institute for Cosmological Physics at the University of Chicago. Z.D. is supported by the NSF Graduate Research Fellowship Program, grant DGE-1144082. R.J.F. is supported in part by NSF grant AST-1518052 and from a fellowship from the Alfred P.\ Sloan Foundation.

Funding for the \DES{} Projects has been provided by the U.S. Department of Energy, the U.S. National Science Foundation, the Ministry of Science and Education of Spain, 
the Science and Technology Facilities Council of the United Kingdom, the Higher Education Funding Council for England, the National Center for Supercomputing 
Applications at the University of Illinois at Urbana-Champaign, the Kavli Institute of Cosmological Physics at the University of Chicago, 
the Center for Cosmology and Astro-Particle Physics at the Ohio State University,
the Mitchell Institute for Fundamental Physics and Astronomy at Texas A\&M University, Financiadora de Estudos e Projetos, 
Funda{\c c}{\~a}o Carlos Chagas Filho de Amparo {\`a} Pesquisa do Estado do Rio de Janeiro, Conselho Nacional de Desenvolvimento Cient{\'i}fico e Tecnol{\'o}gico and 
the Minist{\'e}rio da Ci{\^e}ncia, Tecnologia e Inova{\c c}{\~a}o, the Deutsche Forschungsgemeinschaft and the Collaborating Institutions in the Dark Energy Survey. 

The Collaborating Institutions are Argonne National Laboratory, the University of California at Santa Cruz, the University of Cambridge, Centro de Investigaciones Energ{\'e}ticas, 
Medioambientales y Tecnol{\'o}gicas-Madrid, the University of Chicago, University College London, the \DES{}-Brazil Consortium, the University of Edinburgh, 
the Eidgen{\"o}ssische Technische Hochschule (ETH) Z{\"u}rich, 
Fermi National Accelerator Laboratory, the University of Illinois at Urbana-Champaign, the Institut de Ci{\`e}ncies de l'Espai (IEEC/CSIC), 
the Institut de F{\'i}sica d'Altes Energies, Lawrence Berkeley National Laboratory, the Ludwig-Maximilians Universit{\"a}t M{\"u}nchen and the associated Excellence Cluster Universe, 
the University of Michigan, the National Optical Astronomy Observatory, the University of Nottingham, The Ohio State University, the University of Pennsylvania, the University of Portsmouth, 
SLAC National Accelerator Laboratory, Stanford University, the University of Sussex, Texas A\&M University, and the OzDES Membership Consortium.

The \DES{} data management system is supported by the National Science Foundation under Grant Number AST-1138766.
The \DES{} participants from Spanish institutions are partially supported by MINECO under grants AYA2012-39559, ESP2013-48274, FPA2013-47986, and Centro de Excelencia Severo Ochoa SEV-2012-0234.
Research leading to these results has received funding from the European Research Council under the European Union's Seventh Framework Programme (FP7/2007-2013) including ERC grant agreements 
 240672, 291329, and 306478.
 
This work was supported in part by the Kavli Institute for Cosmological Physics at the University of Chicago through grant NSF PHY-1125897 and an endowment from the Kavli Foundation and its founder Fred Kavli.

\bibliographystyle{apa}
\bibliography{bib}
\end{document}